\documentclass[useAMS,usenatbib]{mn2e}
\usepackage{graphicx}
\usepackage{txfonts}
\usepackage{natbib}

\title[]{Evidence for a change in the radiation mechanism in the hard state
of GRO J1655--40. Hysteresis in the broad-band noise components}
\author[P. Reig et al.]{P. Reig$^{1,2}$\thanks{E-mail: pau@physics.uoc.gr},
     I.E. Papadakis$^{2,1}$ , M.A. Sobolewska$^{3}$, J. Malzac$^{4,5}$\\
$^{1}$IESL, Foundation for Research and Technology, 71110 Heraklion,
Crete, Greece\\
$^{2}$Department of Physics and Institute of Theoretical \& Computational 
Physics, University of Crete,PO Box 2208, GR-710 03, Heraklion, Crete,
Greece \\
$^{3}$ Harvard-Smithsonian Center for Astrophysics, 60, Garden Street,
Cambridge, MA 02138, USA \\
$^{4}$Universit\'e de Toulouse; UPS-OMP; IRAP;  Toulouse, France, \\ 
$^{5}$CNRS; IRAP; 9 Av. colonel Roche, BP 44346, F-31028 Toulouse cedex 4, France
}

\newcommand{\gro}  {GRO J1655--40}
\newcommand{\gx}  {GX 339--4}

\def\simless{\mathbin{\lower 3pt\hbox
     {$\rlap{\raise 5pt\hbox{$\char'074$}}\mathchar"7218$}}}   
\def\simmore{\mathbin{\lower 3pt\hbox
     {$\rlap{\raise 5pt\hbox{$\char'076$}}\mathchar"7218$}}}   

\def\msun{~{\rm M}_\odot}

\begin{document}

\date{Accepted ??. Received ??; in original form ??}

\pagerange{\pageref{firstpage}--\pageref{lastpage}} \pubyear{2013}

\maketitle

\label{firstpage}

\begin{abstract}

We have analysed archival data from the {\it Rossi X-ray Timing Explorer
(RXTE)} to study the aperiodic variability of the black-hole binary GRO
J1655--40 during the hard state of the 2005 outburst.  This work was
motivated by the recent finding of a spectral change in the hard state X-ray
radiation mechanism in black hole binaries. We computed the  0.008-64 Hz
power spectral density during the rise and decay of the 2005
outburst, and we found that they were reasonably well modelled by the sum
of two, broad Lorenztian functions in most cases (plus a narrow QPO)
which correspond to three different variability components. Our aim is to
study the evolution of the  timing properties of the source during the
outburst, by studying the correlation between the  characteristics of the
broad-band noise components in the power spectra and the source 
luminosity. Our results suggest that the whole power spectrum shifts to
high (low) frequencies  as the source luminosity increases (decreases), in
agreement with previous studies of other black hole binaries.
However, we also detect a strong ``hysteresis" pattern in the 
``frequency-luminosity" plots, and show that the ``critical" luminosity
limit, above which the  timing properties of the source change, is
different during the rise and the decay phase of the  outburst. We discuss
the general implications of these results in the context of the truncated
disc model.

\end{abstract}


\section{Introduction}

Black-hole binaries (BHBs) consist of a black hole orbiting a "normal"
companion. By "normal" it is meant that nuclear burning in its interior is
still the main source of energy. BHBs divide up into low-mass and high-mass
systems. Although the two first BHBs discovered, namely Cyg X--1
\citep{webster72} and LMC X--3 \citep{cowley83} were persistent high-mass
BHBs, low-mass transient BHB represent the vast majority of  BHBs.  They
are strong sources of high-energy radiation and show a richness in their
X-ray variability properties. High-mass BHB tend to be persistent sources,
while low-mass BHBs are normally transient sources containing a Roche lobe
overflowing M-A star. Transient BHBs exhibit X-ray outbursts lasting for a
few weeks \citep{remillard06}. During these outbursts the X-ray luminosity
increases by several orders of magnitude from a quiescent level of 
$10^{-6}-10^{-8}$ $L_{\rm Edd}$ \citep{gallo08} to a peak level of $\sim
L_{\rm Edd}$, where $L_{\rm Edd}$ is the Eddington luminosity ($L_{\rm Edd}
\approx 1.3 \times 10^{38} \; (M/\msun)$ erg cm$^{-2}$ s$^{-1}$, for a
black hole of mass $M$). Thus low-mass BHBs are also called ``X-ray novae".

Hard X-rays in BHBs are thought to be produced in the vicinity of the black
hole, in the region between the event horizon and the inner parts of the
accretion disc. The main emission mechanisms that are currently believed to
produce the X-rays in the 2--100 keV band are Compton up-scattering and
blackbody radiation. The spectrum from the accretion disc is the sum of
blackbody components, with increasing temperature and luminosity, as the
disc radius decreases. Typically, the temperature in the innermost part of
the disc is 0.1-0.3 keV. These low-energy photons emitted by the disc gain
energy by means of inverse Compton in a region around the black hole that
is generally called as the ``corona", generating a continuum spectrum with
a power-law shape which extends up to a  few hundred keV. The formation and
heating mechanism, as well as the geometry, of the corona is one of the
unsolved issues in black-hole binary physics. The corona could even be the
base of a mildy relativistic radio jet \citep{markoff05}. Alternatively, 
the hard X-rays detected in the hard state could be generated by inverse
Comptonization in the jet itself \citep[see e.g.][and references
therein]{kylafis08}. 

The monitoring of the evolution of the X-ray spectral and timing parameters
over the course of an outburst reveals significant changes in their
spectral and timing properties. Although these properties change smoothly
as the outburst progresses, most of the phenomenology is better understood
by invoking two main source states: a ``hard" and a ``soft"
state\footnote{For a detailed classification of black-hole states and state
transitions see e.g. \citet{belloni10}.}. The relative contribution of the
above mentioned mechanisms can broadly explain the spectral properties of
the sources in the two states: whenever the accretion disc dominates, the
power-law continuum is soft, and a broad, thermal, blackbody-like spectrum
is also observed; if emission from the corona dominates, then only the
Comptonised power-law "tail" shows up in the spectrum (with slopes which
are flatter than those detected in the soft state). The hard state
generally appears at the beginning and end of the outbursts, while the soft
state is observed at the peak of the outburst. 

\begin{table}
\begin{center}
\caption{Summary of the RXTE observations.}
\label{xobs}
\begin{tabular}{ccc}
\hline \hline \noalign{\smallskip}
Proposal &MJD		&On-source	      \\
ID	 &		&time (ks)	     \\
\hline \noalign{\smallskip}
P90058	&53424.01--53431.17	&15.2	     \\
P90428	&53426.04--53432.79	&83.5	   \\
P90704	&53439.74--53439.61	&6.3	   \\
P91404	&53433.90--53436.40     &17.1	   \\
P91702	&53436.72--53660.97     &212.2	   \\
P91704	&53637.17--53637.50     &19.8	   \\
\hline \hline \noalign{\smallskip}
\end{tabular}
\end{center}
\end{table}

\gro\ is one of the best studied BHBs. It is one of the most firmly
established black hole candidates and its astrophysical parameters are
fairly well known. \gro\ was discovered by BATSE onboard the Compton Gamma
Ray Observatory in July 1994 \citep{zhang94}. It is located at a distance
of $3.2\pm0.2$ kpc \citep[][but see \citealt{foellmi06}]{hjellming95} and
contains a black hole with a mass of $6.3\pm$0.3 $\msun$
\citep{remillard06} and spin $a=0.78\pm0.10$ \citep{gierlinski01}. The
optical counterpart is an F3 IV-F6 IV star with a mass of 2.3 $\msun$ in a
2.6 day orbit. The inner disk is viewed at an inclination of 70$^\circ$
\citep{orosz97}. 

Past timing studies of \gro\ have focused mainly on the properties of
quasi-periodic oscillations (QPO) that are detected in its power spectral
density (PSDs).  \citet{shaposhnikov07} analysed data from the early stages
of the  2005 outburst and found that the low-frequency QPO strongly
correlates with  the spectral parameters. They concluded that the region
where the QPO originates moves inward as the outburst progressed. The
correlation between spectral and timing parameters can also be used to
estimate the mass of the black hole \citep{shaposhnikov09}. The smooth
day-to-day variation of the QPO frequency during the 2005 outburst led
\citet{chakrabarti08} to suggest that an oscillating shock, which sweeps
inward through the disk in the rising phase and outward in declining phase,
is responsible for the QPO. The explanation of the shock oscillations was
preferred to that one that assumes that QPO are generated by orbiting blobs
at the inner edge of the accretion disk because the QPO is a stable and
lasting feature, which has to survive for weeks \citep{debnath08}. 
\citet{motta12} made a detailed analysis of the QPOs in \gro\ under
the $ABC$ classification scheme \citep[see e.g.][]{casella05}. These authors
reported the {\em simultaneous} detection of type-B and type C QPO during
the peak of the outbursts, in the so-called ultraluminous state, indicating
that these two types of QPO gave a different physical origin. 
\citet{motta12} also studied the properties of the peaked noise component
detected in the ultra-luminous state.

Recently, \citet[][S11 hereafter, see also \citealt{wu08}]{sobolewska11}
performed an X-ray {\em spectral} analysis and found that the ``power-law
spectral index -- luminosity" relation of \gro\ (and \gx) changes when  the
source luminosity is greater than $L_{crit,S11} =0.006\pm 0.002$ of the
Eddington limit.  Motivated by this result,  we studied  in this work the
broad-band noise components that dominate the X--ray aperiodic variability
of \gro, using {\it RXTE} data   collected during the 2005 outburst of the
source. We are mainly interested in investigating  the variation of their
parameters with the source's luminosity and spectral shape,  with the hope
that this X--ray {\em timing} analysis may shed light on the X--ray source
geometry,  the accretion mode as well as the radiation mechanisms that
operate when the source is in the hard state.

\begin{figure}
\includegraphics[width=8cm]{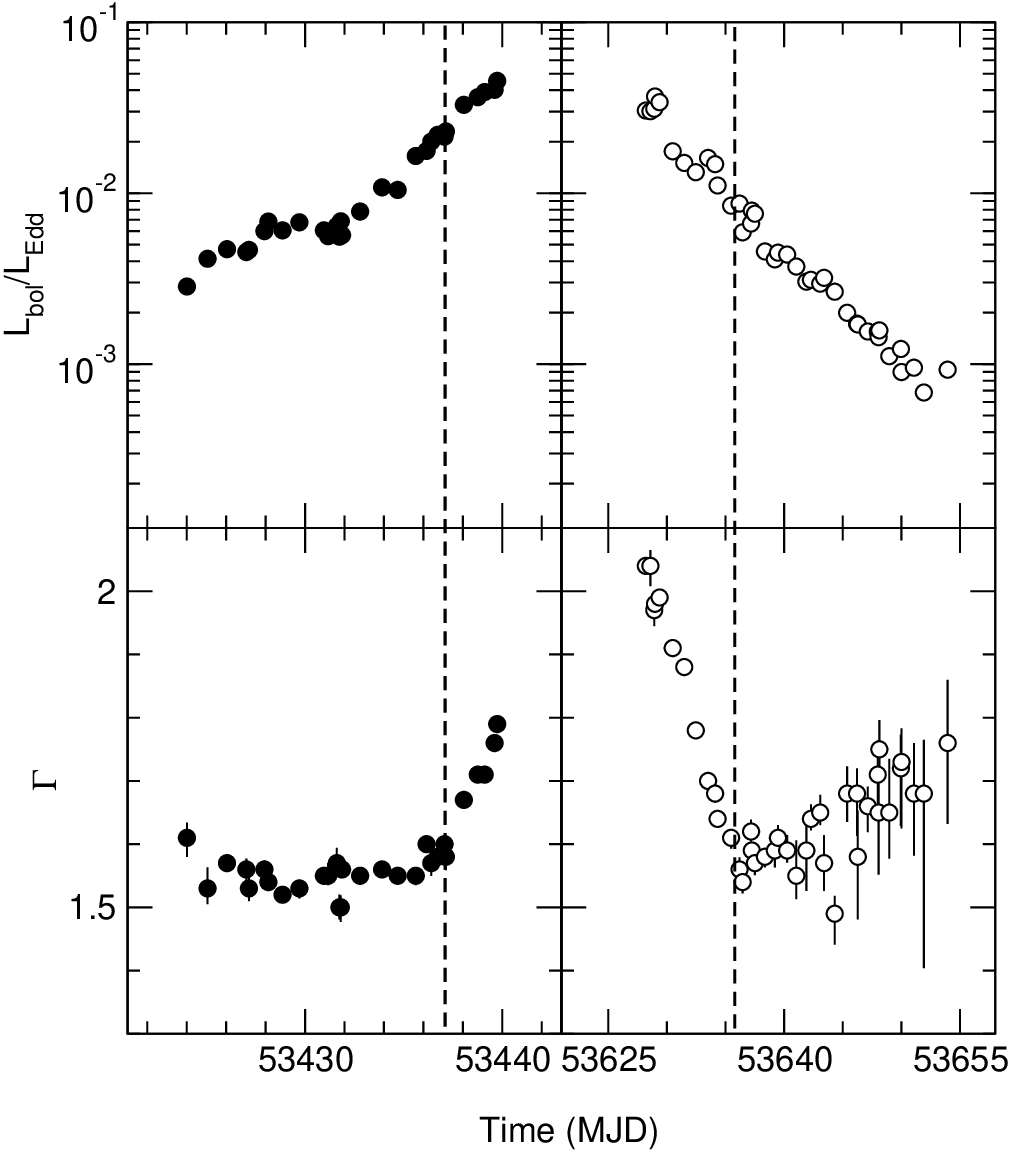} 
\caption[]{X-ray luminosity (upper panel) and power-law photon index (lower
panel) as a function of time. Filled and opened symbols
represent the rise and decay of the outburst, respectively.
The vertical dashed lines indicate the time of the first
(left) and last (right) appearance of $L_0$ (see Sect.~\ref{psd}).}
\label{lc}
\end{figure}

\begin{figure*}
\begin{tabular}{cc}
\includegraphics[width=8cm]{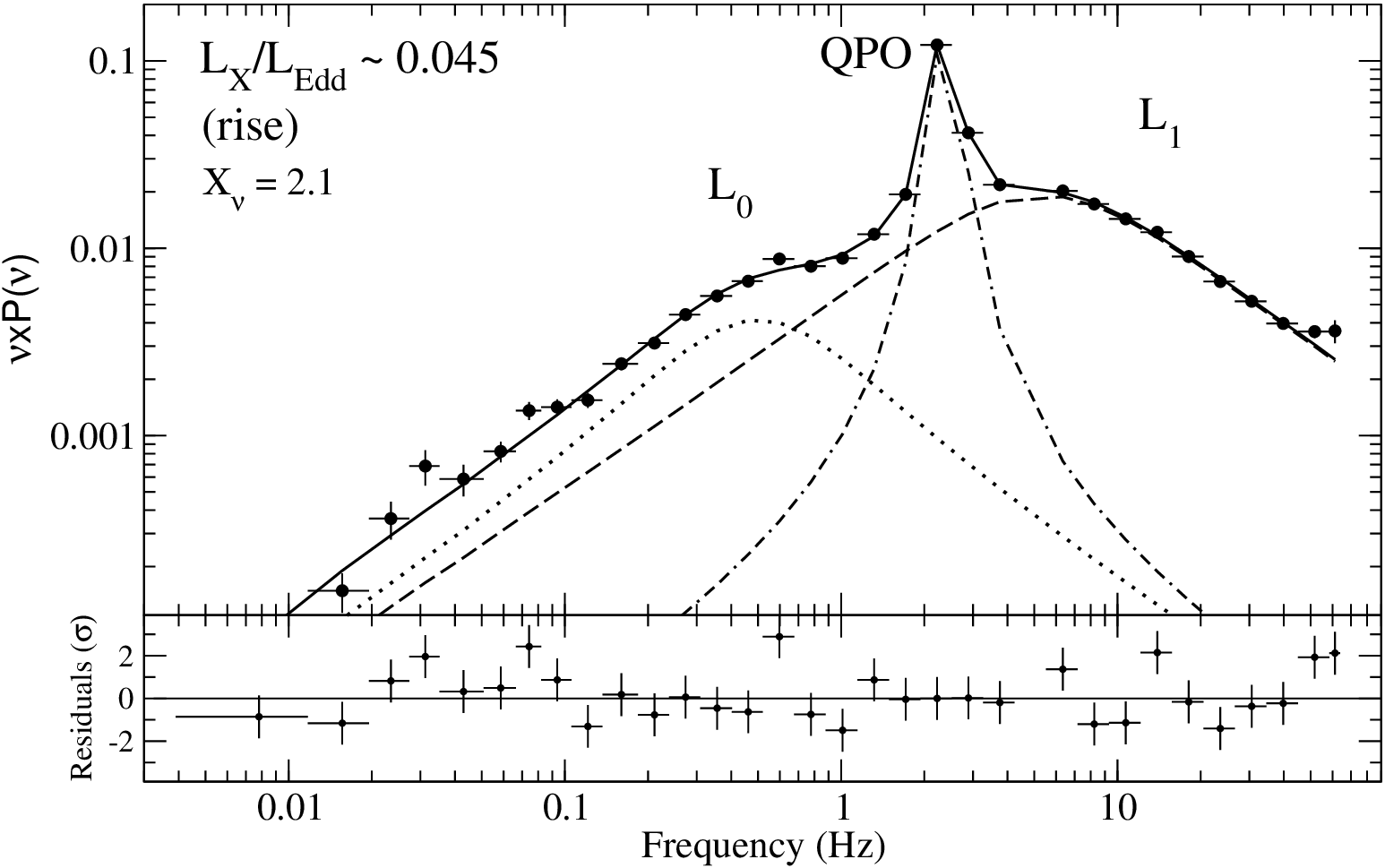} &
\includegraphics[width=8cm]{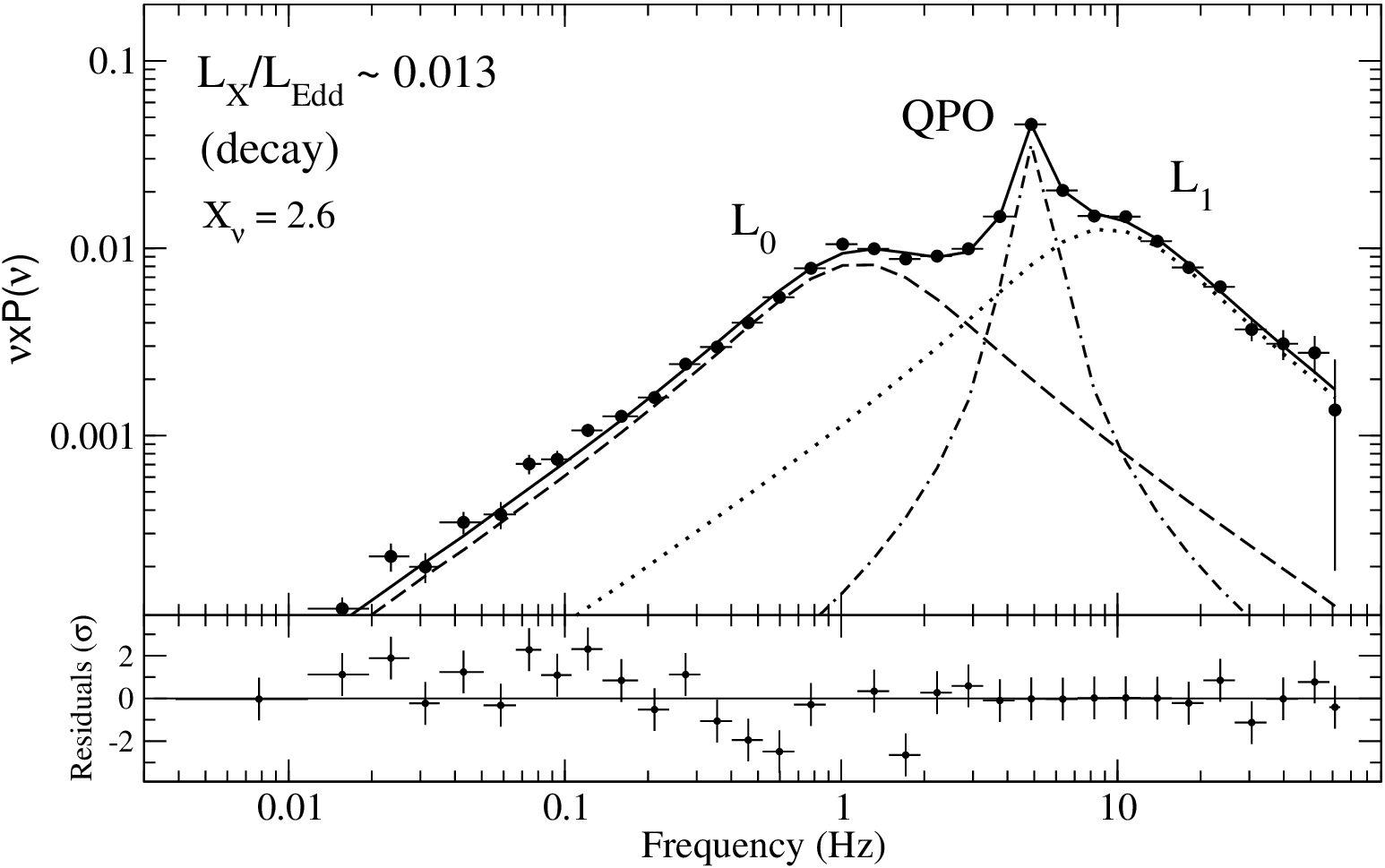} \\
\includegraphics[width=8cm]{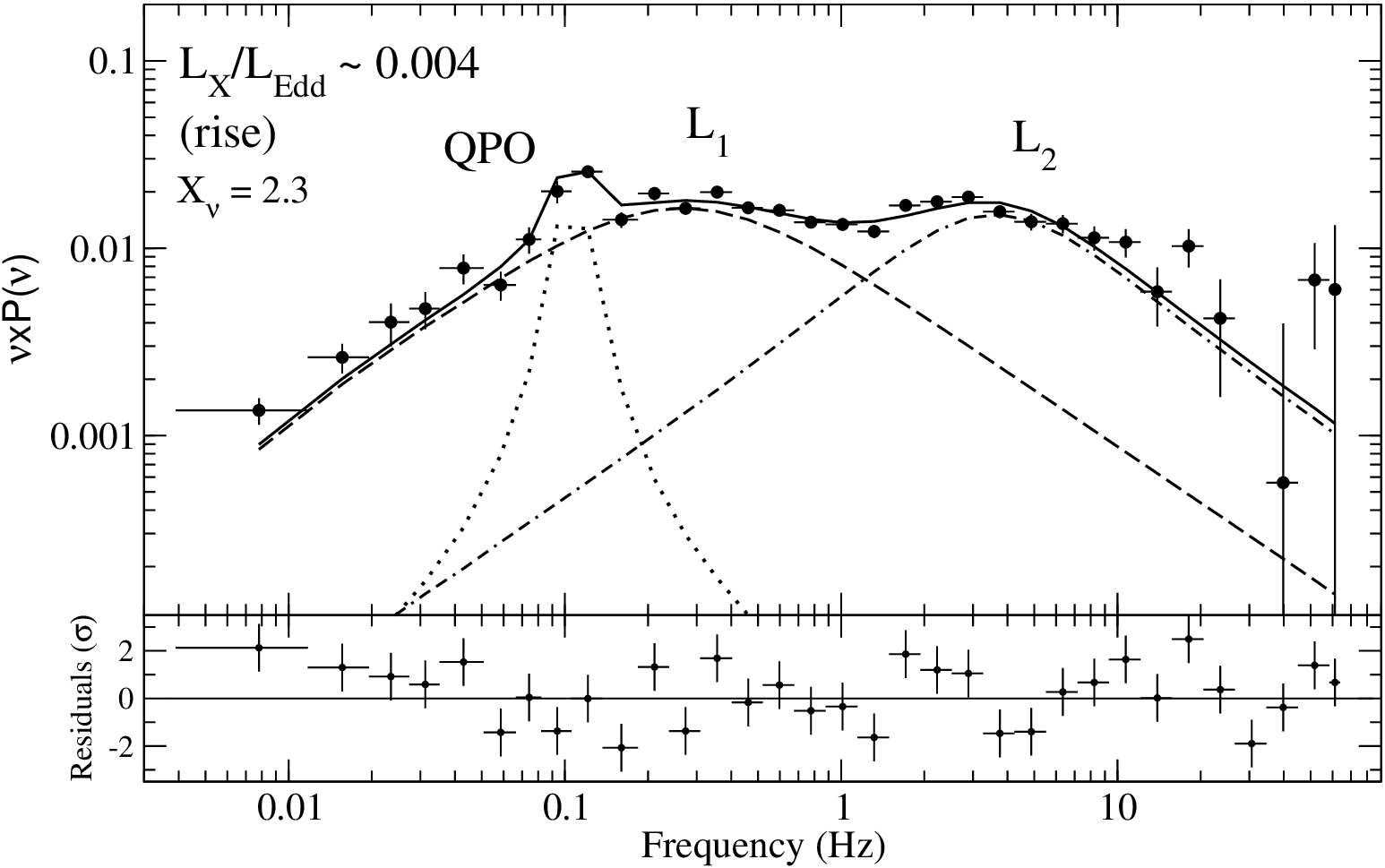} &
\includegraphics[width=8cm]{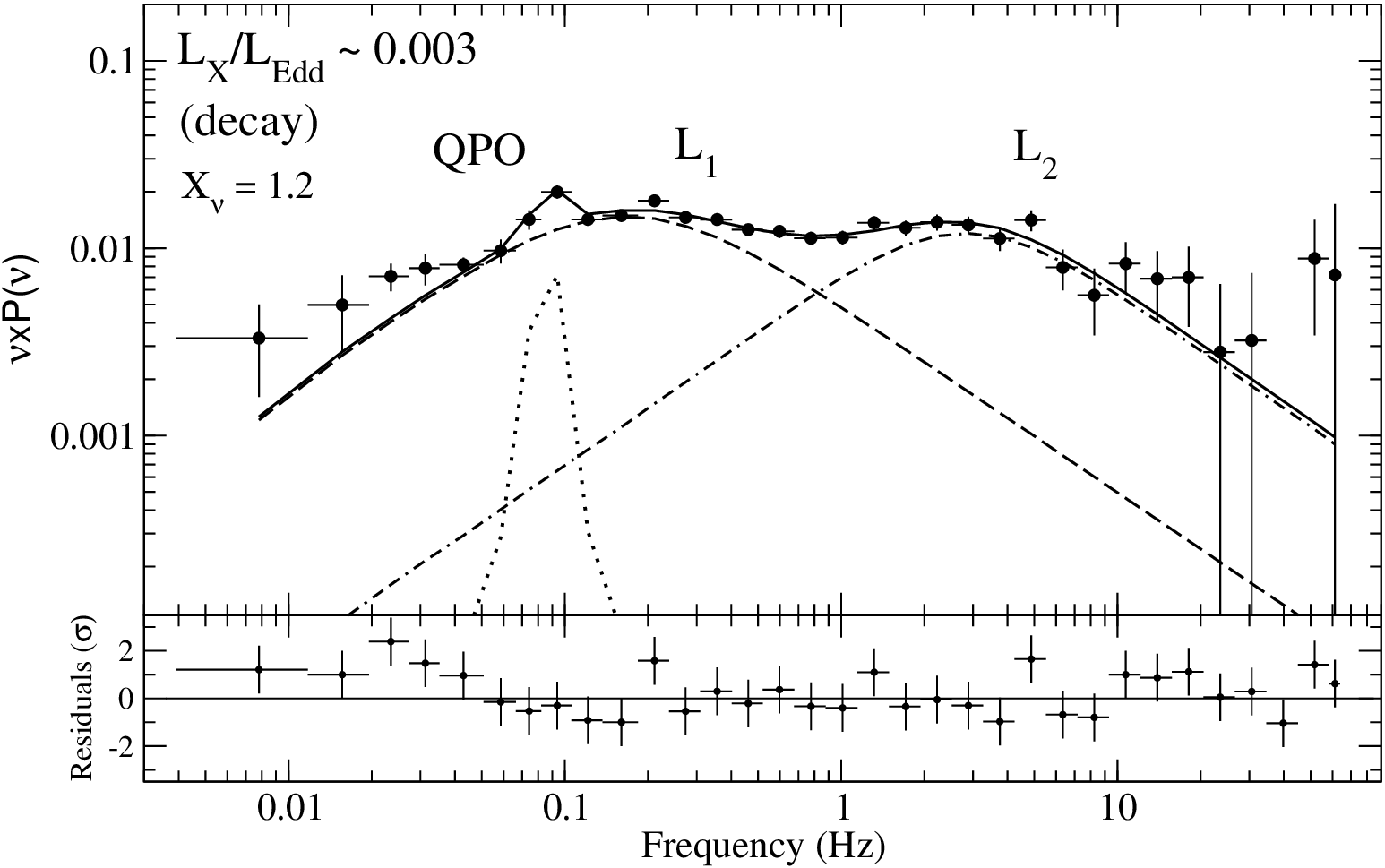} \\

\end{tabular}
\caption[]{Representative power spectra of \gro\ during the rise and decay
phases of the outburst. }
\label{psdplot}
\end{figure*}

\section{Observations and data analysis}

\subsection{The {\it RXTE} observations}

We analysed {\it RXTE} data of  \gro\ during the 2005
outburst. Since we wish to investigate fast time X-ray variability
($\sim$0.1-100 s) we only used data from the Proportional Counter (PCA).
The PCA consists of five Proportional Counter Units (PCUs) with a total
collecting area of $\sim$6250 cm$^2$. It operates in the 2--60 keV range and
has a nominal energy resolution of 18\% at 6 keV \citep{jahoda96}. The data
were retrieved from the {\it RXTE} archived and correspond to the
observation programs given in Table ~\ref{xobs}.

We selected the data with the same criteria as in  S11, namely,  we
discarded all observations  with a photon index of $\Gamma \simmore 2$.
The source was in the canonical hard state in all observations of the
rise phase of the outburst. The decay phase include hard-intermediate and
hard state observations \citep{motta12}. The photon index values resulted
from the best fit to the PCA+HEXTE 3--200 keV spectra as described in
\citet{sobolewska09}. We only considered the {\it RXTE} data up to MJD
53654, which include  light curves with an average count rate larger than
22.5 counts s$^{-1}$ in the 2--13 keV band, because the PSD estimation  is
less reliable at lower count rates.  The total number of observations which
satisfy this criterion are 69, representing a total on-source exposure time
of 326 ks. 

Fig.~\ref{lc} shows the X-ray luminosity (from 0.01 to
1000 keV) in Eddington units, plotted as a function of time (on MJD) for
the selected {\it RXTE} observations. Because the majority of energy in
BHBs is radiated in the X-ray band, we consider the X-ray luminosity to be
representative of the bolometric flux, hence we denote with `$L_{\rm
bol}/L_{\rm Edd}$' the ratio of the X-ray luminosity to the Eddington
limit. We used the $L_{\rm bol}/L_{\rm Edd}$ values of S11, which were
based on the best, {\tt DISKBB + EQPAIR} model fits to the spectra of the
source (see S11 for details). The upper panel in Fig.~1 indicates that the 
data we use in this work cover well the rise and the decay of the 2005
X-ray outburst of \gro. In the bottom panel of the same figure we plot
$\Gamma$ as a function of time. This plot indicates the spectral evolution
of the source during the rising and decaying phases of the outburst, which
was studied in detail by S11.

\begin{table*}
\begin{center}
\caption{Best-fit parameters of the $L_0$, $L_1$ and $L_2$ components for
the rise phase of the outburst.}
\label{rise}
\begin{tabular}{@{~~~~}c@{~~~~}c@{~~~~}c@{~~~~}c@{~~~~}c@{~~~~}c@{~~~~}c@{~~~~}c@{~~~~}c@{~~~~}c@{~~~~}c@{~~~~}c@{~~~~}c}
\hline \hline \noalign{\smallskip}
MJD	&$L_{\rm bol}/L_{\rm Edd}$ &$RMS$(\%)	&$Q_0$	&$\nu_0$	&$rms_0$	&$Q_1$	&$\nu_1$	&$rms_1$	&$Q_2$	&$\nu_2$	&$rms_2$	&$\chi^2_{r}$\\
	&			   &(0.01-10Hz)	&	&(Hz)		&		&	&(Hz)		&	\\
\hline \noalign{\smallskip}
53424.009 & 0.0028 & 30.4 & --- & --- & --- & 0.10 & $0.094^{+0.029}_{-0.008}$ & $0.307^{+0.039}_{-0.025}$ & 0.00 & $2.485^{+0.593}_{-0.587}$ & $0.300^{+0.022}_{-0.124}$ & 1.30  \\ 
53425.055 & 0.0041 & 30.5 & --- & --- & --- & 0.00 & $0.147^{+0.023}_{-0.012}$ & $0.311^{+0.012}_{-0.020}$ & 0.00 & $2.540^{+0.701}_{-0.691}$ & $0.298^{+0.011}_{-0.011}$ & 2.08  \\ 
53426.040 & 0.0047 & 30.6 & --- & --- & --- & 0.03 & $0.149^{+0.009}_{-0.008}$ & $0.313^{+0.006}_{-0.004}$ & 0.00 & $3.070^{+0.213}_{-0.202}$ & $0.319^{+0.005}_{-0.005}$ & 1.46  \\ 
53427.022 & 0.0045 & 29.8 & --- & --- & --- & 0.00 & $0.264^{+0.021}_{-0.024}$ & $0.321^{+0.008}_{-0.009}$ & 0.26 & $3.596^{+0.513}_{-0.448}$ & $0.241^{+0.010}_{-0.006}$ & 2.35  \\ 
53427.155 & 0.0047 & 28.9 & --- & --- & --- & 0.00 & $0.242^{+0.020}_{-0.015}$ & $0.316^{+0.007}_{-0.018}$ & 0.25 & $3.851^{+0.317}_{-0.306}$ & $0.244^{+0.011}_{-0.006}$ & 1.30  \\ 
53427.942 & 0.0060 & 29.5 & --- & --- & --- & 0.05 & $0.231^{+0.015}_{-0.016}$ & $0.307^{+0.007}_{-0.007}$ & 0.12 & $3.697^{+0.297}_{-0.219}$ & $0.274^{+0.006}_{-0.005}$ & 1.28  \\ 
53428.138 & 0.0068 & 29.5 & --- & --- & --- & 0.01 & $0.186^{+0.018}_{-0.014}$ & $0.311^{+0.013}_{-0.032}$ & 0.11 & $3.012^{+0.267}_{-0.273}$ & $0.278^{+0.038}_{-0.019}$ & 0.88  \\ 
53428.859 & 0.0061 & 29.8 & --- & --- & --- & 0.00 & $0.226^{+0.010}_{-0.009}$ & $0.335^{+0.005}_{-0.009}$ & 0.19 & $3.722^{+0.182}_{-0.173}$ & $0.255^{+0.022}_{-0.012}$ & 1.46  \\ 
53429.712 & 0.0067 & 29.4 & --- & --- & --- & 0.00 & $0.271^{+0.025}_{-0.022}$ & $0.330^{+0.004}_{-0.005}$ & 0.21 & $3.878^{+0.315}_{-0.340}$ & $0.245^{+0.016}_{-0.010}$ & 1.52  \\ 
53430.958 & 0.0061 & 30.2 & --- & --- & --- & 0.05 & $0.171^{+0.015}_{-0.013}$ & $0.327^{+0.024}_{-0.019}$ & 0.00 & $3.220^{+3.610}_{-0.432}$ & $0.297^{+0.010}_{-0.012}$ & 1.18  \\ 
53431.173 & 0.0056 & 27.8 & --- & --- & --- & 0.00 & $0.176^{+0.018}_{-0.015}$ & $0.291^{+0.009}_{-0.013}$ & 0.02 & $3.067^{+0.123}_{-0.005}$ & $0.283^{+0.011}_{-0.008}$ & 1.98  \\ 
53431.612 & 0.0064 & 29.1 & --- & --- & --- & 0.00 & $0.288^{+0.027}_{-0.021}$ & $0.288^{+0.009}_{-0.009}$ & 0.15 & $3.945^{+0.447}_{-0.406}$ & $0.255^{+0.009}_{-0.036}$ & 1.49  \\ 
53431.743 & 0.0056 & 29.1 & --- & --- & --- & 0.00 & $0.229^{+0.041}_{-0.030}$ & $0.314^{+0.006}_{-0.008}$ & 0.16 & $3.543^{+0.329}_{-0.312}$ & $0.259^{+0.026}_{-0.005}$ & 2.31  \\ 
53431.810 & 0.0068 & 28.1 & --- & --- & --- & 0.09 & $0.298^{+0.054}_{-0.051}$ & $0.287^{+0.025}_{-0.028}$ & 0.17 & $4.011^{+0.737}_{-0.706}$ & $0.249^{+0.053}_{-0.032}$ & 1.89  \\ 
53431.875 & 0.0057 & 29.0 & --- & --- & --- & 0.00 & $0.241^{+0.022}_{-0.024}$ & $0.319^{+0.004}_{-0.007}$ & 0.16 & $4.101^{+0.361}_{-0.384}$ & $0.258^{+0.007}_{-0.017}$ & 2.41  \\ 
53432.793 & 0.0078 & 29.2 & --- & --- & --- & 0.00 & $0.329^{+0.019}_{-0.019}$ & $0.338^{+0.004}_{-0.007}$ & 0.29 & $4.617^{+0.439}_{-0.468}$ & $0.226^{+0.014}_{-0.012}$ & 2.69  \\ 
53433.905 & 0.0108 & 28.4 & --- & --- & --- & 0.00 & $0.510^{+0.018}_{-0.020}$ & $0.335^{+0.004}_{-0.004}$ & 0.23 & $5.764^{+0.388}_{-0.312}$ & $0.223^{+0.007}_{-0.003}$ & 1.21  \\ 
53434.694 & 0.0104 & 28.4 & --- & --- & --- & 0.00 & $0.645^{+0.046}_{-0.048}$ & $0.329^{+0.006}_{-0.008}$ & 0.00 & $6.250^{+0.835}_{-0.700}$ & $0.261^{+0.008}_{-0.022}$ & 2.53  \\ 
53435.612 & 0.0165 & 28.3 & --- & --- & --- & 0.06 & $0.795^{+0.073}_{-0.076}$ & $0.336^{+0.005}_{-0.013}$ & 0.18 & $8.469^{+1.211}_{-1.284}$ & $0.205^{+0.028}_{-0.023}$ & 1.16  \\ 
53436.159 & 0.0177 & 28.9 & --- & --- & --- & 0.10 & $0.749^{+0.049}_{-0.029}$ & $0.311^{+0.013}_{-0.005}$ & 0.10 & $6.874^{+0.539}_{-0.324}$ & $0.243^{+0.020}_{-0.004}$ & 2.46  \\ 
53436.398 & 0.0200 & 29.0 & --- & --- & --- & 0.00 & $1.045^{+0.055}_{-0.052}$ & $0.362^{+0.006}_{-0.017}$ & 0.36 & $10.247^{+1.060}_{-1.193}$ & $0.174^{+0.021}_{-0.006}$ & 3.59  \\ 
53436.725 & 0.0220 & 29.3 & --- & --- & --- & 0.01 & $1.055^{+0.052}_{-0.061}$ & $0.358^{+0.003}_{-0.005}$ & 0.22 & $9.356^{+0.713}_{-0.763}$ & $0.196^{+0.019}_{-0.003}$ & 2.21  \\ 
53437.072 & 0.0214 & 29.5 & --- & --- & --- & 0.05 & $1.050^{+0.076}_{-0.079}$ & $0.345^{+0.019}_{-0.018}$ & 0.31 & $9.371^{+0.929}_{-1.043}$ & $0.187^{+0.022}_{-0.018}$ & 2.66  \\ 
53437.142 & 0.0230 & 30.3 & 0.00 & $0.012^{+0.064}_{-0.025}$ & $0.113^{+0.110}_{-0.039}$ & 0.12 & $1.140^{+0.176}_{-0.102}$ & $0.330^{+0.023}_{-0.080}$ & 0.26 & $10.121^{+1.638}_{-1.587}$ & $0.193^{+0.024}_{-0.019}$ & 2.25  \\ 
53438.054 & 0.0329 & 30.0 & 0.21 & $0.289^{+0.108}_{-0.085}$ & $0.139^{+0.046}_{-0.031}$ & 0.41 & $1.978^{+0.189}_{-0.219}$ & $0.32^{+0.030}_{-0.019}$ & 0.00 & $10.477^{+3.713}_{-2.273}$ & $0.227^{+0.013}_{-0.021}$ & 1.30  \\ 
53438.757 & 0.0364 & 30.6 & 0.67 & $0.301^{+0.020}_{-0.024}$ & $0.094^{+0.004}_{-0.004}$ & 0.00 & $3.335^{+0.120}_{-0.090}$ & $0.359^{+0.000}_{-0.000}$ & --- & --- & --- & 2.44  \\ 
53439.107 & 0.0390 & 30.7 & 0.75 & $0.313^{+0.031}_{-0.033}$ & $0.081^{+0.005}_{-0.006}$ & 0.04 & $3.385^{+0.067}_{-0.061}$ & $0.366^{+0.023}_{-0.002}$ & --- & --- & --- & 3.11  \\ 
53439.610 & 0.0402 & 30.4 & 0.52 & $0.372^{+0.037}_{-0.030}$ & $0.098^{+0.005}_{-0.003}$ & 0.14 & $4.130^{+0.051}_{-0.105}$ & $0.330^{+0.002}_{-0.002}$ & --- & --- & --- & 3.53  \\ 
53439.741 & 0.0453 & 29.5 & 0.38 & $0.498^{+0.058}_{-0.054}$ & $0.114^{+0.011}_{-0.010}$ & 0.15 & $5.265^{+0.210}_{-0.213}$ & $0.301^{+0.012}_{-0.011}$ & --- & --- & --- & 2.09  \\ 
\hline \hline \noalign{\smallskip}
\end{tabular}
\end{center}
\end{table*}

\begin{table*}
\begin{center}
\caption{Best-fit parameters of the $L_0$, $L_1$ and $L_2$ components for
the decay phase of the outburst.}
\label{decay}
\begin{tabular}{@{~~~~}c@{~~~~}c@{~~~~}c@{~~~~}c@{~~~~}c@{~~~~}c@{~~~~}c@{~~~~}c@{~~~~}c@{~~~~}c@{~~~~}c@{~~~~}c@{~~~~}c}
\hline \hline \noalign{\smallskip}
MJD	&$L_{\rm bol}/L_{\rm Edd}$ &$RMS$(\%)	&$Q_0$	&$\nu_0$	&$rms_0$	&$Q_1$	&$\nu_1$	&$rms_1$	&$Q_2$	&$\nu_2$	&$rms_2$	&$\chi^2_{r}$\\
	&			   &(0.01-10Hz)	&	&(Hz)		&		&	&(Hz)		&	\\
\hline \noalign{\smallskip}
53628.196 & 0.0304 &  7.6 & 0.08 & $6.256^{+1.353}_{-1.310}$ & $0.126^{+0.006}_{-0.008}$ & ---  & --- & --- & ---  & --- & --- & 1.66  \\ 
53628.590 & 0.0301 &  9.4 & 0.00 & $5.698^{+0.840}_{-0.720}$ & $0.151^{+0.010}_{-0.009}$ & ---  & --- & --- & ---  & --- & --- & 2.19  \\ 
53628.917 & 0.0311 & 11.8 & 0.14 & $4.130^{+0.667}_{-0.317}$ & $0.154^{+0.015}_{-0.013}$ & ---  & --- & --- & ---  & --- & --- & 1.50  \\ 
53628.983 & 0.0368 & 11.1 & 0.31 & $3.187^{+0.289}_{-0.280}$ & $0.134^{+0.002}_{-0.003}$ & ---  & --- & --- & ---  & --- & --- & 1.69  \\ 
53629.376 & 0.0341 & 12.4 & 0.25 & $3.221^{+0.455}_{-0.370}$ & $0.147^{+0.004}_{-0.029}$ & ---  & --- & --- & ---  & --- & --- & 1.66  \\ 
53630.490 & 0.0176 & 14.6 & 0.34 & $2.392^{+0.236}_{-0.250}$ & $0.135^{+0.010}_{-0.008}$ & 0.43 & $13.399^{+4.832}_{-4.852}$ & $0.129^{+0.006}_{-0.006}$ & 0.00 & --- & --- & 1.71  \\ 
53631.474 & 0.0150 & 17.3 & 0.38 & $1.915^{+0.041}_{-0.096}$ & $0.159^{+0.004}_{-0.007}$ & 0.96 & $13.821^{+1.397}_{-2.404}$ & $0.123^{+0.032}_{-0.012}$ & 0.00 & --- & --- & 3.04  \\ 
53632.456 & 0.0133 & 21.5 & 0.42 & $1.115^{+0.066}_{-0.060}$ & $0.154^{+0.009}_{-0.008}$ & 0.34 & $8.483^{+0.906}_{-0.650}$ & $0.202^{+0.013}_{-0.014}$ & 0.00 & --- & --- & 2.60  \\ 
53633.504 & 0.0161 & 22.5 & 0.50 & $0.534^{+0.068}_{-0.063}$ & $0.096^{+0.011}_{-0.009}$ & 0.11 & $4.743^{+0.329}_{-0.307}$ & $0.280^{+0.019}_{-0.017}$ & 0.00 & --- & --- & 1.56  \\ 
53634.109 & 0.0148 & 23.1 & 0.69 & $0.416^{+0.059}_{-0.044}$ & $0.086^{+0.013}_{-0.010}$ & 0.00 & $4.612^{+0.466}_{-0.445}$ & $0.308^{+0.007}_{-0.006}$ & 0.00 & --- & --- & 1.85  \\ 
53634.315 & 0.0111 & 23.0 & 0.27 & $0.480^{+0.202}_{-0.123}$ & $0.119^{+0.036}_{-0.018}$ & 0.00 & $4.304^{+0.650}_{-0.620}$ & $0.286^{+0.008}_{-0.010}$ & 0.00 & --- & --- & 1.51  \\ 
53635.470 & 0.0084 & 23.8 & --- & --- & --- & 0.00 & $0.969^{+0.036}_{-0.035}$ & $0.303^{+0.004}_{-0.005}$ & 0.09 & $7.861^{+6.620}_{-2.206}$ & $0.184^{+0.009}_{-0.007}$ & 1.70  \\ 
53636.190 & 0.0087 & 23.6 & --- & --- & --- & 0.00 & $0.804^{+0.053}_{-0.042}$ & $0.291^{+0.006}_{-0.007}$ & 0.00 & $10.468^{+5.700}_{-3.366}$ & $0.219^{+0.012}_{-0.009}$ & 1.48  \\ 
53636.452 & 0.0059 & 23.4 & --- & --- & --- & 0.00 & $0.675^{+0.080}_{-0.042}$ & $0.295^{+0.007}_{-0.007}$ & 0.00 & $7.027^{+7.014}_{-0.548}$ & $0.200^{+0.014}_{-0.014}$ & 2.18  \\ 
53637.175 & 0.0066 & 24.7 & --- & --- & --- & 0.00 & $0.424^{+0.048}_{-0.050}$ & $0.280^{+0.011}_{-0.015}$ & 0.00 & $4.530^{+1.100}_{-0.980}$ & $0.244^{+0.012}_{-0.020}$ & 1.68  \\ 
53637.238 & 0.0079 & 24.5 & --- & --- & --- & 0.00 & $0.433^{+0.025}_{-0.023}$ & $0.287^{+0.004}_{-0.004}$ & 0.00 & $5.300^{+0.700}_{-0.600}$ & $0.231^{+0.005}_{-0.005}$ & 4.13  \\ 
53637.500 & 0.0076 & 24.0 & --- & --- & --- & 0.00 & $0.412^{+0.042}_{-0.035}$ & $0.273^{+0.007}_{-0.011}$ & 0.00 & $5.550^{+1.130}_{-0.910}$ & $0.244^{+0.009}_{-0.030}$ & 2.72  \\ 
53638.353 & 0.0046 & 24.4 & --- & --- & --- & 0.00 & $0.299^{+0.033}_{-0.028}$ & $0.266^{+0.007}_{-0.006}$ & 0.00 & $3.685^{+0.970}_{-0.650}$ & $0.246^{+0.009}_{-0.012}$ & 1.83  \\ 
53639.204 & 0.0041 & 25.9 & --- & --- & --- & 0.00 & $0.310^{+0.043}_{-0.037}$ & $0.299^{+0.009}_{-0.027}$ & 0.00 & $5.500^{+1.790}_{-1.350}$ & $0.256^{+0.014}_{-0.013}$ & 1.64  \\ 
53639.466 & 0.0045 & 25.3 & --- & --- & --- & 0.00 & $0.214^{+0.036}_{-0.035}$ & $0.254^{+0.009}_{-0.036}$ & 0.00 & $2.690^{+0.447}_{-0.533}$ & $0.276^{+0.010}_{-0.012}$ & 1.52  \\ 
53640.252 & 0.0044 & 25.5 & --- & --- & --- & 0.00 & $0.192^{+0.021}_{-0.016}$ & $0.286^{+0.009}_{-0.021}$ & 0.00 & $3.707^{+0.810}_{-0.645}$ & $0.259^{+0.013}_{-0.013}$ & 2.82  \\ 
53641.038 & 0.0037 & 25.8 & --- & --- & --- & 0.00 & $0.225^{+0.031}_{-0.029}$ & $0.302^{+0.011}_{-0.037}$ & 0.15 & $3.020^{+0.984}_{-0.850}$ & $0.210^{+0.053}_{-0.033}$ & 1.53  \\ 
53641.889 & 0.0030 & 27.5 & --- & --- & --- & 0.00 & $0.170^{+0.028}_{-0.023}$ & $0.304^{+0.010}_{-0.026}$ & 0.00 & $3.790^{+0.920}_{-0.730}$ & $0.270^{+0.016}_{-0.034}$ & 1.93  \\ 
53642.283 & 0.0031 & 27.1 & --- & --- & --- & 0.01 & $0.146^{+0.011}_{-0.012}$ & $0.310^{+0.009}_{-0.023}$ & 0.00 & $3.085^{+0.600}_{-0.600}$ & $0.260^{+0.011}_{-0.026}$ & 1.64  \\ 
53643.070 & 0.0030 & 27.7 & --- & --- & --- & 0.00 & $0.146^{+0.017}_{-0.015}$ & $0.306^{+0.007}_{-0.009}$ & 0.00 & $2.765^{+0.505}_{-0.420}$ & $0.275^{+0.011}_{-0.009}$ & 1.23  \\ 
53643.402 & 0.0032 & 26.8 & --- & --- & --- & 0.00 & $0.114^{+0.026}_{-0.025}$ & $0.321^{+0.012}_{-0.013}$ & 0.00 & $3.707^{+1.150}_{-0.850}$ & $0.259^{+0.013}_{-0.013}$ & 1.07  \\ 
53644.313 & 0.0027 & 28.2 & --- & --- & --- & 0.00 & $0.102^{+0.015}_{-0.016}$ & $0.271^{+0.039}_{-0.025}$ & 0.00 & $3.024^{+1.124}_{-0.873}$ & $0.269^{+0.021}_{-0.020}$ & 1.39  \\ 
53645.370 & 0.0020 & 29.0 & --- & --- & --- & 0.09 & $0.105^{+0.015}_{-0.010}$ & $0.293^{+0.028}_{-0.024}$ & 0.00 & $3.049^{+0.705}_{-0.725}$ & $0.302^{+0.015}_{-0.015}$ & 1.14  \\ 
53646.213 & 0.0017 & 27.4 & --- & --- & --- & 0.18 & $0.085^{+0.016}_{-0.016}$ & $0.264^{+0.019}_{-0.020}$ & 0.00 & $2.395^{+0.930}_{-0.700}$ & $0.291^{+0.023}_{-0.052}$ & 1.63  \\ 
53646.279 & 0.0017 & 27.7 & --- & --- & --- & 0.00 & $0.101^{+0.015}_{-0.013}$ & $0.316^{+0.015}_{-0.019}$ & 0.00 & $2.993^{+1.471}_{-0.937}$ & $0.264^{+0.024}_{-0.025}$ & 1.45  \\ 
53647.133 & 0.0016 & 26.4 & --- & --- & --- & 0.13 & $0.076^{+0.015}_{-0.011}$ & $0.264^{+0.008}_{-0.017}$ & 0.08 & $1.251^{+0.305}_{-0.305}$ & $0.242^{+0.020}_{-0.030}$ & 1.60  \\ 
53647.982 & 0.0016 & 29.0 & --- & --- & --- & 0.20 & $0.068^{+0.013}_{-0.011}$ & $0.285^{+0.048}_{-0.041}$ & 0.00 & $1.048^{+0.304}_{-0.463}$ & $0.198^{+0.122}_{-0.041}$ & 1.30  \\ 
53648.048 & 0.0014 & 28.3 & --- & --- & --- & 0.00 & $0.077^{+0.015}_{-0.011}$ & $0.333^{+0.016}_{-0.016}$ & 0.00 & $1.930^{+1.885}_{-0.836}$ & $0.259^{+0.049}_{-0.042}$ & 1.48  \\ 
53648.113 & 0.0016 & 26.6 & --- & --- & --- & 0.03 & $0.057^{+0.012}_{-0.009}$ & $0.264^{+0.029}_{-0.015}$ & 0.00 & $1.880^{+1.540}_{-0.840}$ & $0.281^{+0.042}_{-0.041}$ & 1.76  \\ 
53648.965 & 0.0011 & 29.0 & --- & --- & --- & 0.39 & $0.064^{+0.012}_{-0.010}$ & $0.319^{+0.018}_{-0.018}$ & 0.00 & $6.687^{+2.185}_{-2.652}$ & $0.336^{+0.082}_{-0.068}$ & 1.25  \\ 
53649.951 & 0.0012 & 27.4 & --- & --- & --- & 0.17 & $0.048^{+0.019}_{-0.016}$ & $0.269^{+0.048}_{-0.038}$ & 0.00 & $1.015^{+0.540}_{-0.651}$ & $0.275^{+0.026}_{-0.074}$ & 2.19  \\ 
53650.013 & 0.0009 & 26.3 & --- & --- & --- & 0.00 & $0.046^{+0.008}_{-0.000}$ & $0.311^{+0.015}_{-0.038}$ & 0.00 & $1.035^{+0.670}_{-0.430}$ & $0.236^{+0.030}_{-0.060}$ & 1.39  \\ 
53651.061 & 0.0010 & 26.2 & --- & --- & --- & 0.15 & $0.041^{+0.010}_{-0.010}$ & $0.264^{+0.049}_{-0.037}$ & 0.00 & $0.990^{+0.850}_{-0.461}$ & $0.263^{+0.039}_{-0.065}$ & 0.92  \\ 
53651.912 & 0.0007 & 27.8 & --- & --- & --- & 0.00 & $0.038^{+0.006}_{-0.006}$ & $0.308^{+0.017}_{-0.017}$ & 0.00 & $1.022^{+0.792}_{-0.426}$ & $0.235^{+0.040}_{-0.036}$ & 1.60  \\ 
53653.943 & 0.0009 & 25.8 & --- & --- & --- & 0.00 & $0.046^{+0.013}_{-0.014}$ & $0.288^{+0.023}_{-0.038}$ & 0.00 & $2.720^{+3.440}_{-2.560}$ & $0.271^{+0.147}_{-0.107}$ & 1.05  \\ 
53657.156 & 0.0005 & 28.8 & --- & --- & --- & 0.00 & $0.027^{+0.007}_{-0.011}$ & $0.253^{+0.031}_{-0.029}$ & 0.00 & $10.000^{+7.}_{-7.}$      & $0.323^{+0.072}_{-0.089}$ & 1.75  \\ 
\hline \hline \noalign{\smallskip}
\end{tabular}
\end{center}
\end{table*}

\subsection{Computation and fitting of the power spectra}

Power spectra (PSD) were computed using PCA data in the energy range 2--13
keV. Light curves with a time  resolution of $2^{-7}$ s were divided into
128-s segments and a Fast Fourier Transform was computed for each segment.
Thus the frequency range sampled by the power spectra was $0.008-64$ Hz.
The final power spectrum was computed as the average of all power spectra
obtained for each segment. These averaged power spectra were
logarithmically rebinned in frequency and corrected for dead time effects
according to the prescriptions given in \citet{nowak99}. Power spectra were
normalized in a way that the integral is equal to
the squared fractional rms amplitude, according to the so-called
rms-normalization. 

Usually, models that include Lorentzian profiles fit well the broad-band
PSD of BHBs in their hard state \citep{belloni02,pottschmidt03,axelsson05}.
We therefore constructed a conservative model with as few Lorenzian
profiles as possible, to fit the PSDs in our sample. The
Lorentzian functions are characterized by their normalization, $R$ (which
determines the total rms amplitude of the Lorentzian), their resonance
frequency, $\nu_c$, and its quality factor, $Q$, which is defined as
$Q=\nu/FWHM$ ($FWHM$ is the full frequency width of the Lorentzian at half
of its maximum value). Traditionally, profiles with $Q>2$ are referred to
as ``narrow" and are indicative of quasi-periodic oscillations. On the
other hand, profiles with $Q<2$ are known as peaked noise and correspond to
``broad-band" components in the PSDs. Instead of $\nu_c$, we used the peak
frequency of the Lorenztian (i.e. the frequency where the contribution of
the Lorentzian to the total rms is maximized), which is defined as
$\nu_{\rm max}=(\nu_c^2+(FWHM/2)^2)^{1/2}$ in the  $\nu \times P_{\nu}$
representation of the PSD. All the Lorentzian parameters were left free to
vary during the fitting procedure.

We found that most of the PSDs were well fitted by the sum of three
Lorentzians. One of them was ``narrow"  and corresponds to the QPO feature
that has been studied in the past. The other two turned out to be
``broad".  As we argue in the next section, our results indicate that the
broad Lorentzian components correspond to {\it three}  different types of
noise, that we shall refer to as $L_i$ with $i=0,1,2$.  Only  two PSDs
(those which correspond to the  observations taken on MJD=53437.1417  and
53438.0537) are best fitted by a model which includes the three broad
Lorentzian  components, that is, the three types of noise (plus the QPO). 
In all other PSDs, $L_0$ and $L_2$ do not appear together. In five
observations (MJD 53628--53629) only one component (plus the QPO) was
enough to fit the PSD.

Figure~\ref{psdplot} shows four representative power spectra at low and high
luminosity for the rising and decaying phases of the outburst. The plots of
the power spectra are shown in the $\nu \times P_{\nu}$ representation,
where each power is multiplied by the corresponding frequency
\citep{belloni02}.

In Tables~\ref{rise} and \ref{decay} we list the best-fit values of the
Q-parameter ($Q_i$), maximum frequency ($\nu_{i}$), and the fractional
amplitude of variability $rms_i$ (normalisation of the Lorentzian), for
each one of the $L_0$, $L_1$, and $L_2$ broad Lorentzian components. Also
listed is the  overall $RMS$, estimated by the integral of the overall PSD
over the observed frequency range. Note that Tables~\ref{rise} and
\ref{decay} do not include the  $L_{\rm QPO}$ best-fit parameter vaules. 

The best-fit $\chi^2_{r}$ indicate that the quality of the fit is not
statistically acceptable in quite a few cases.  The residuals do not always
imply a good fit due to the presence of statistically significant
``wiggles" in the data. In some cases, these wiggles are indicative of
sharp features, and in other cases of weaker and broader features. However,
we did not attempt to model further these features, as our aim is not to
provide the best possible description to the PSDs of the source, but rather
to provide a reliable characterization of the broad PSD shape at all
luminosities and investigate its evolution. As the plots of the ``bad fit
quality" PSDs in Fig.~\ref{psdplot} show, even in this case, two main
Lorentzian profiles, plus the QPO, can describe accurately the main
characteristics of the broad band noise components in the PSD of the
source. 

\section{PSD model fitting results}

\subsection{The evolution of the Lorentzian peak frequencies and rms amplitude}
\label{psd}

\begin{figure}
\includegraphics[width=8cm]{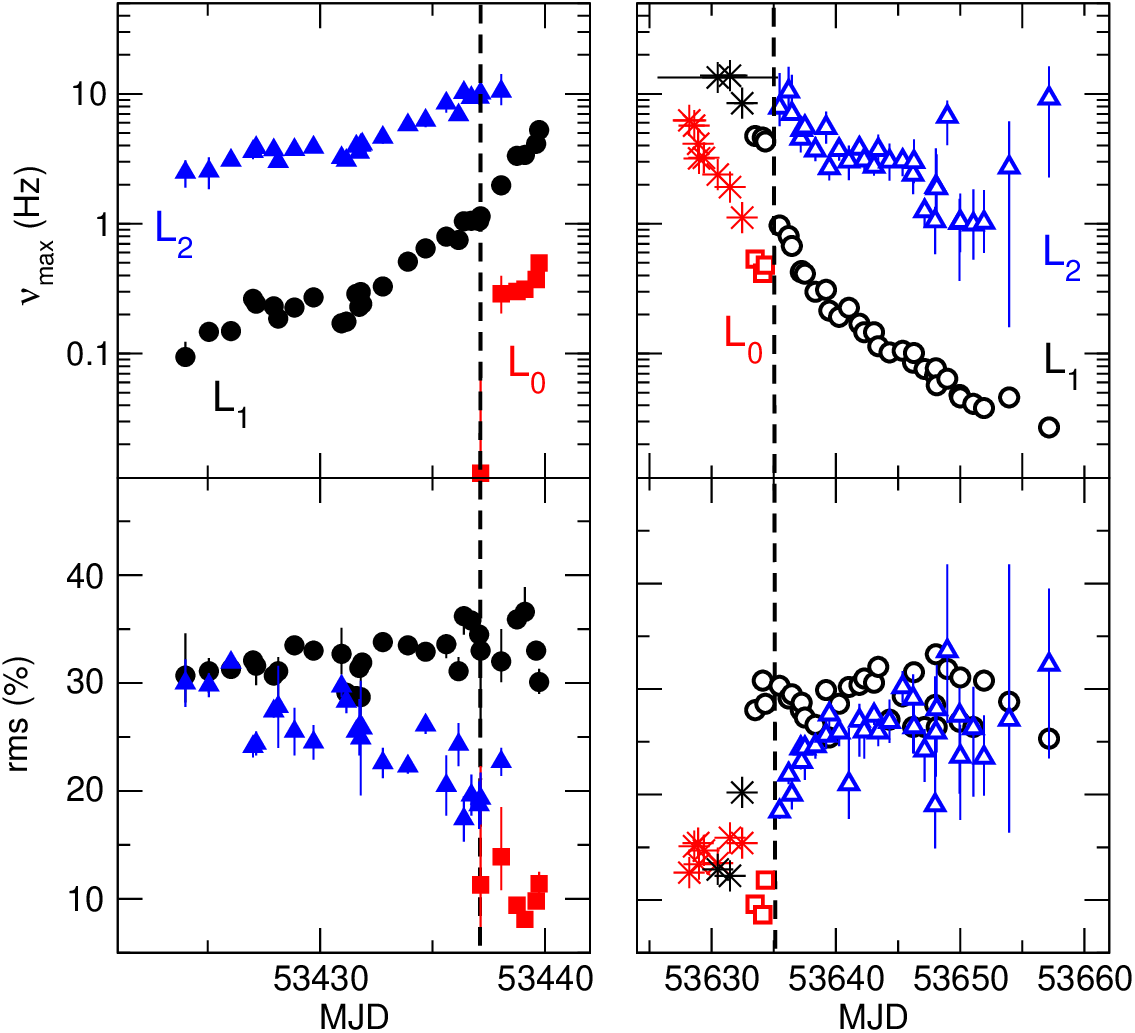} 
\caption[]{Time evolution of the peak frequency and rms amplitude  (top and
bottom panel, resepctively) of the noise components during the rise  and
decay phases of the outburst (left and right panel, respectively). Red
squares, black circles and blue triangles indicate the $L_0$, $L_1$, and
$L_2$ best-fit parameter  values, while filled (open) symbols correspond to
the rise (decay) parts of the  outburst. Stars represent the
hard-intermediate state observations. The vertical dashed lines indicate
the time of the first (left) and last (right) appearance of $L_0$, which
corresponds to a luminosity of 0.03 and 0.01 $L_{\rm Edd}$, respectively.}
\label{results1}
\end{figure}

The time evolution of the best-fit $L_i$  peak frequencies and rms during
the rise  (filled symbols) and decay phase (open symbols) of the outburst
is shown on the  top and bottom panels  of  Fig.~\ref{results1},
respectively.  The star-like symbols in the right panels (decay)
represent the hard-intermediate state observations. At the beginning of
the outburst,  the PSDs are well described by the sum of two Lorentzians,
whose peak frequencies  increase with the source luminosity (left top panel
of Fig.~\ref{results1}). We denote with  $L_1$ and $L_2$ the lower and
higher frequency Lorentzians, respectively. The dashed line  indicates the
time at which a third Lorentzian appears in the PSD, with a peak frequency 
$\sim 10$ times lower than $\nu_{\rm max,1}$. We denote this low frequency
component as $L_0$. The source luminosity is $\sim 0.03$ of the Eddington
limit at this stage.  At higher luminosities, $L_2$ is not detected any
more, and the PSDs are well fitted by the  sum of $L_0$ and $L_1$.

The situation is more complicated during the decay phase of the outburst.
At the beginning,  the PSD is well described by a single Lorentzian.  We
identify this component as the $L_0$  Lorentzian. At this point, the
source finds itself in the hard-intermediate state. When $L_{\rm
bol}/L_{\rm Edd}$ decreases to $\sim 0.02$, a second  Lorentzian appears
with a higher peak frequency, which we identify as the $L_1$. However,  at
MJD$\sim 53635$, when $L_{\rm bol}\sim 0.01L_{\rm Edd}$ (indicated by the 
dashed vertical line in  the  right panels  of Fig.~\ref{results1}), the
$L_1$ peak Lorentzian frequencies decreases abruptly by  a factor of $\sim
4$, while a new component with peak frequency two times higher than
$\nu_{\rm max,1}$ appears. We believe that at this point $L_2$ shows up
again and $L_0$ is not  detected any more. At all later times, the PSDs are
well fitted by the sum of $L_1$ and $L_2$. The PSDs in which both
$L_0$ and $L_1$ are present may be identified with the transition from the
hard-intermediate state to the hard state. The abrupt change marks the
beginning of the hard state.

During the rise phase of the outburst, rms$_1$ increases slightly from
$\sim 30$ to $\sim 35$\%,  while rms$_2$ decreases monotonically with
increasing source flux (from $\sim 30$ to $\sim 20$\% ;  left bottom panel
in Fig.~\ref{results1}). During the decay phase, rms$_1$ and rms$_2$ are
less than  $\sim 20$\% when $L_1$ and $L_2$ appear for the first time.  The
$L_1$ and $L_2$ rms amplitudes  increase as the source flux decreases until
they reach the level of $\sim 30$\% and $\sim25$\%, respectively.  At later
times, they remain roughly constant at this level. On the contrary,  
rms$_0$ does not show any significant variations during the source flux
evolution.  It remains constant at a level of $\sim 10-15$\% during both
the rise and decay phase of the outburst.  The higher values of rms$_0$
correspond to the hard-intermediate state.

Our results indicate that $L_0$  is a broad Lorentzian ($Q\simless 0.7$),
which peaks in the range   of 0.1--6 Hz. It is only detected in 6 out of 29
PSDs during the rise (when $L>0.03L_{\rm Edd}$)  and 11 out of 41 PSDs
during the decay (when $L>0.01L_{\rm Edd}$). $L_1$ is almost always 
present, except for the first five PSDs during the decay phase of the
outburst, when the source flux is high,  and rms$_1<10$\%.  Its
characteristic frequency varies in the range 0.05--10 Hz. This component 
is consistent with being a zero-centred Lorentzian ($Q\sim 0$) in most
PSDs. $L_2$ is detected only when  $L<0.03L_{\rm Edd}$, in the rise phase,
and $L<0.01L_{\rm Edd}$, during the decay.  $Q_2$ is always smaller than 
$\sim$0.3, and most of the time is consistent with zero.

\begin{figure}
\includegraphics[width=8cm]{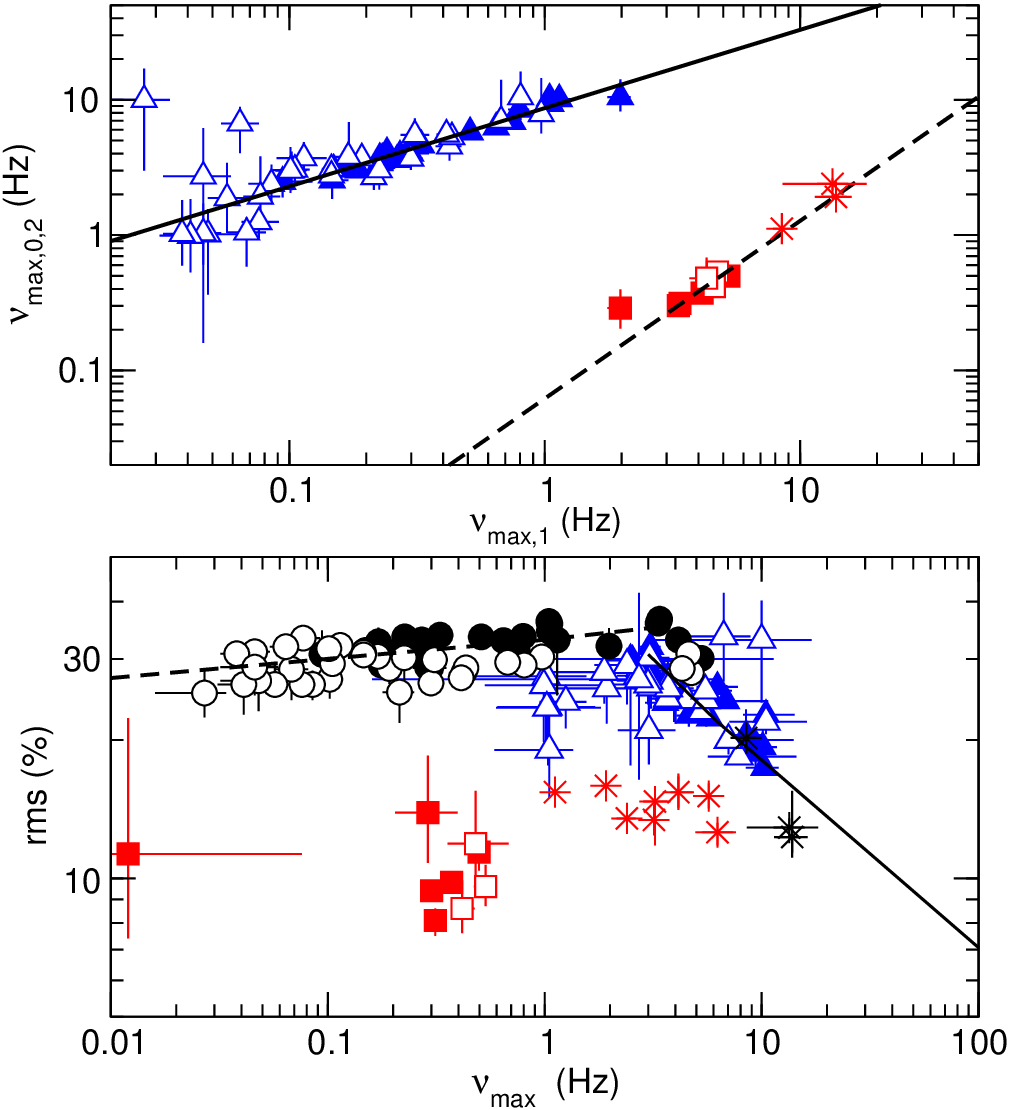} 
\caption[]{Top panel: the $L_0$ and $L_2$ peak frequencies plotted as a 
function of $\nu_{\rm max,1}$ (squares and triangles, respectively) and 
best-fit lines.  Bottom panel: the Lorentzian rms plotted as a function of
the respective peak frequencies (red squares, black circles and blue
triangles for the $L_0$, $L_1$  and $L_0$ components, respectively). Filled
and open symbols indicate the  rise and decay parts of the outburst in both
panels. Stars represent the hard-intermediate state observations. The
dashed line represents the best fit to the $L_1$ points below 1 Hz, while
the solid line  fits the $L_2$ points above 3 Hz.}
\label{results2}
\end{figure}

The top panel in Fig.~\ref{results2} shows the ``$\nu_{\rm max,0}$ {\it vs}
$\nu_{\rm max,1}$" and ``$\nu_{\rm max,2}$ {\it vs} $\nu_{\rm max,1}$"
plots (squares and triangles, respectively;  filled and open symbols
indicate the rise and decay phase of the outburst).   Clearly, the peak
frequencies evolve in a highly correlated way during the outburst. We
fitted both data sets in the top panel of  Fig.~\ref{results2} with a power
law model. The solid and dashed lines in this panel indicate the best-fit
results to the ``$\nu_{\rm max,2}$ vs $\nu_{\rm max,1}$" and ``$\nu_{\rm
max,0}$ vs $\nu_{\rm max,1}$"  data sets, respectively. The fit was done in
the log-log space), using the {\tt fitxy} subroutine of \citet{press92},
which takes into account the errors on both variables (we  assumed that 
the $\nu_{\rm max}$ errors are symmetric and equal to the mean of the
best-fit positive and negative  confidence limits listed in Table 2). The
best-fit results are as follows:  log($\nu_{\rm
max,2})=0.94(\pm0.02)+0.58(\pm 0.03)$log($\nu_{\rm max,1})$ ($\chi^2=46.8$
for 53 degrees of freedom -  dof) and log($\nu_{\rm
max,0})=-1.21(\pm0.07)+1.3(\pm 0.1)$log$(\nu_{\rm max,1}) $ ($\chi^2=6.1$
for 9 dof).  The best fit results indicate that the peak frequencies do not
increase (or decrease) with the same  rate. The  $\nu_{\rm max,2}\propto
\nu_{\rm max,1}^{0.6}$ and $\nu_{\rm max,0}\propto \nu_{\rm max,1}^{1.3}$
relations suggest that  $\nu_{\rm max,1}$ evolves  faster than  $\nu_{\rm
max,2}$ and slower than $\nu_{\rm max,0}$, during the rise and decay phases
of the outburst.

The bottom panel in Fig.~\ref{results2} shows the ``rms {\it vs} peak
frequency" plots for the  three Lorentzians. We observe a bimodal behaviour
in the case of $L_1$ and $L_2$. At low characteristic frequencies, 
rms$_1$  increases slowly  with increasing frequency,  until $\nu_{\rm
max,1}\sim 1-3$ Hz.   The dashed line in the same panel indicates the best
fit linear model to the  (rms$_1$,$\nu_{\rm max,1}$) data (in the log-log
space) when $\nu_{\rm max}< 1$ Hz. The best-fit results  indicate that a
relation of the form: rms$_1\propto \nu_{\rm max,1}^{0.042\pm0.005}$, fits
the  data reasonably well. Although the slope is shallow, it is
nevertheless significantly  different from zero. This relation appears to
fit the data well up to $\nu_{\rm max,1}\sim 3$ Hz.  At higher frequencies,
rms$_1$ anti-correlates with $\nu_{\rm max,1}$, both for the rise and  the
decay data. This behaviour is also seen in Cyg X--1
\citep[][see also \citealt{klein-wolt08}]{pottschmidt03,axelsson05}.  

We observe a similar behaviour between rms$_2$ and $\nu_{\rm max,2}$.  The
solid  line in the bottom panel of Fig.~\ref{results2} indicates the
best-fit line to all the  (rms$_2$,$\nu_{\rm max,2}$) data at frequencies
greater than 3 Hz. The  results indicate that the best-fit relation of the
form:  rms$_2\propto \nu_{\rm max,2}^{-0.41\pm0.03}$ describes well the 
``rms$_2 - \nu_{\rm max,2}$" anti-correlation at frequencies above 3 Hz. At
lower  frequencies, there may be a positive correlation between rms$_2$ and
$\nu_{\rm max,2}$,  (as is the case with $L_1$) but the number of data
points is not large enough to draw  a positive conclusion.

On the contrary, rms$_0$ is roughly constant at around $\sim 10-15$\%,
irrespective of  the source flux. Perhaps it is  slightly higher during the
decay phase, but the $L_0$ data points in the bottom panel of 
Fig.~\ref{results2} do not suggest any correlation between rms$_0$ and
$\nu_{\rm max,0}$.

\subsection{The ``peak frequency -- luminosity" relation}

\begin{figure}
\includegraphics[width=8cm]{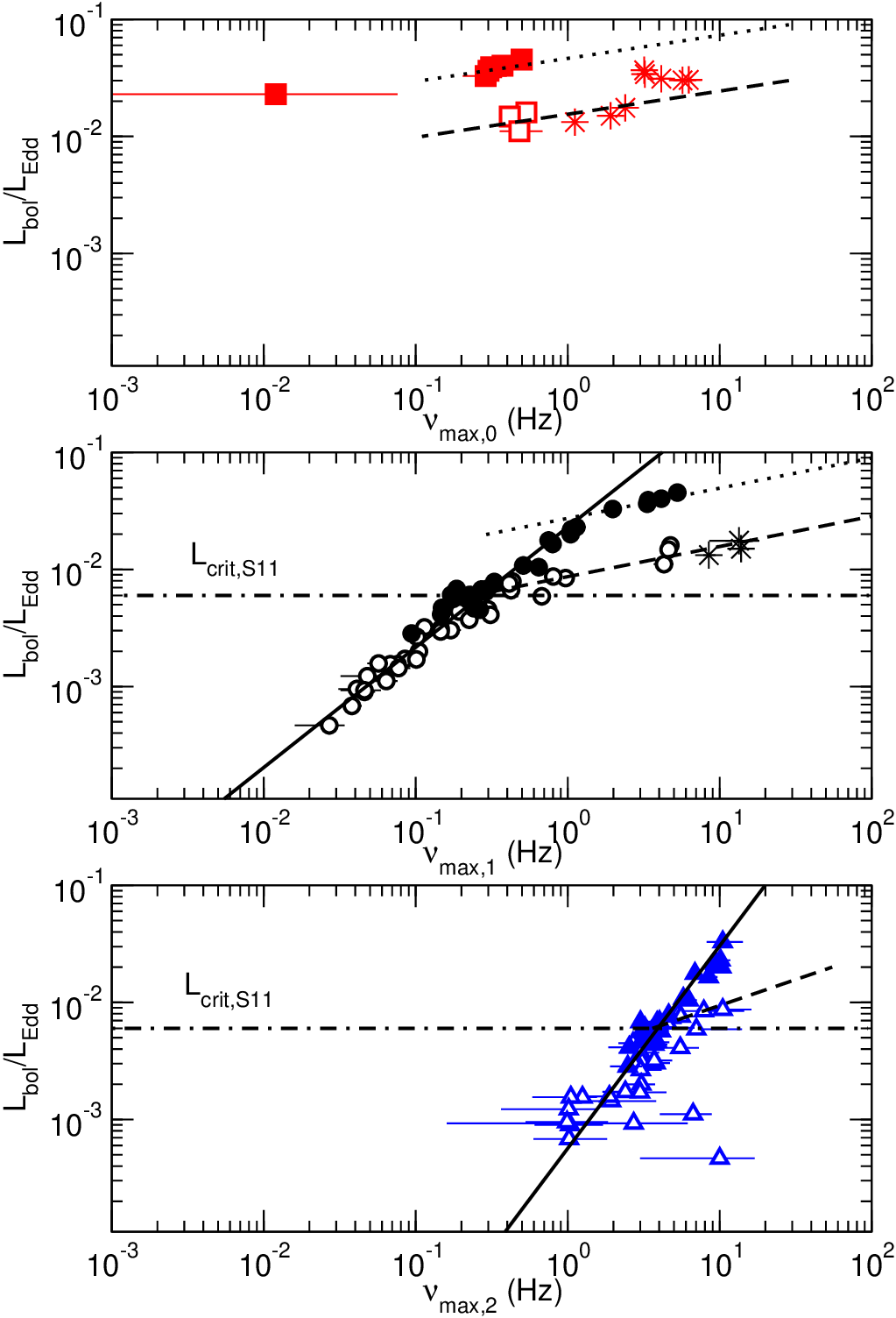} 
\caption[]{The ``peak frequency--luminosity" relation for $L_0$, $L_1$ and
$L_2$  (top, middle and lower panels, respectively). Filled and open
symbols indicate the data sets for the rise and decay outburst  phase,
respectively. Stars (decay) represent the hard-intermediate state
observations. The horizontal dot-dashed line in the middle and bottom
panels  indicates the critical luminosity level of S11, $L_{crit,S11}$.}
\label{fluxnumax} 
\end{figure}

Given the Lorentzian peak frequencies ($\nu_{\rm max}$) evolution during
the outburst (shown in Fig.~\ref{results1}), we expect that $\nu_{\rm max}$
will correlate with the source flux as well. Fig.~\ref{fluxnumax} shows the
relation between $\nu_{\rm max}$ and  $L_{\rm bol}/L_{\rm Edd}$.  In these
plots, we chose to plot  $L_{\rm bol}/L_{\rm Edd}$ in the $y-$axis, similar
to the ``hardness--intensity"  diagrams that are frequently used in the
study of the spectral  evolution of the outbursts in BHBs. Clearly,  the
frequencies correlate positively with source flux.  Since the peak
Lorentzian frequencies are also well correlated with each other (see
Fig.~\ref{results2}),   it appears that the whole PSD shifts to higher
frequencies with increasing flux, and then  shifts back to lower
frequencies, as the flux decreases. 

A hysteresis effect is clearly seen in the $L_0$ and $L_1$ plots. In the
case of $L_0$ (upper panel in  Fig.~\ref{fluxnumax}), at a given luminosity
the peak frequencies during the decay phase (open symbols) are
systematically higher than the respective  frequencies during the rise
phase of the outburst  (filled symbols). The same effect is observed in the
case of $L_1$ (middle panel in  Fig.~\ref{fluxnumax}).  When the
source luminosity is greater than $L_{crit,S11}$  (this ``critical"
luminosity level  is indicated by the horizontal dot-dashed line in the
middle panel of Fig.~\ref{fluxnumax}) the ``$\nu_{\rm max,1}-L_{\rm
bol}/L_{\rm Edd}$" relation during the   decay phase of the outburst is
different from the ``peak frequency -- luminosity"   in the rise phase.  As
a result,  $\nu_{\rm max,1,decay}$ is always greater than $\nu_{\rm
max,1,rise}$ when  $L_{\rm bol}/L_{\rm Edd}>L_{crit,S11}$. At lower
luminosities,  the ``$\nu_{\rm max,1}-L_{\rm bol}/L_{\rm Edd}$" relation is
very similar for both the rise and  the decay data sets.

During the decay phase, the characteristic frequency of $L_1$ displays a
gap between 1 and 3 Hz (middle panel of Fig.~\ref{fluxnumax}). This gap
corresponds to the ``jump" that appears in the $\nu_{\rm max,1}$ evolution
when $L_0$ disappers and $L_2$ appears (upper right panel in
Fig.~\ref{results1}). It is interesting to note that a similar
discontinuity has been reported for the QPO frequency in \gro\ and GX
339--4 only during the decline phase of an outburst (but not during the rise
phase) with no corresponding discontinuity in the count rate
\citep{chakrabarti08,nandi12}.  

To quantify the relation between $\nu_{\rm max,1}$ and $L_{\rm bol}/L_{\rm
Edd}$, we first fitted  the (rise and decay) data below $L_{crit,S11}$ with
a linear function (in the log-log space) of the form: log($\nu_{\rm
max,1})=a_1+\beta_1 \times$log$(L_{\rm bol}/L_{\rm Edd})$. The best-fit
parameter values are: $\beta_1=0.98\pm 0.04$ and $a_1=1.6\pm0.1$. The solid
line in the middle panel of Fig.~\ref{fluxnumax} indicates the best-fit
model to the data. Strictly speaking, the best-fit model is not
statistically accepted  ($\chi^2=187$ for 33 dof). However, as
Fig.~\ref{fluxnumax} shows, the best fit line describes  rather well the
``$\nu_{\rm max,1}-L_{\rm bol}/L_{\rm Edd}$" relation at luminosities lower
than  $L_{crit,S11}$. In fact, this relation appears to also fit well the
[$\nu_{\rm max,1}, (L_{\rm bol}/L_{\rm Edd})$] data  in the rise phase up
to $L\sim 0.03L_{\rm Edd}$. At this luminosity level, $\nu_{\rm max,1}\sim 2$
Hz.  According to the ``rms$_1 - \nu_{\rm max,1}$" relation plotted  in the
bottom panel of Fig~\ref{results2}, at higher frequencies (and hence
luminosities) rms$_1$ starts to decrease. 

The dashed line in the same panel indicates the best-fit of the same linear
model to the  [log($\nu_{\rm max,1}), \log(L_{\rm bol}/L_{\rm Edd}$)] data
during the decay phase, when the source luminosity  is greater than
$L_{crit,S11}$.  The best fit results are: $\beta_1=3.9\pm 0.2$ and
$a_1=8.0\pm0.4$ ($\chi^2=73.5$ for 10 dof). Interestingly, the same model
line (with the normalization increased by a factor of $\sim 3$) represents
well the  ``$\nu_{\rm max,1}-L_{\rm bol}/L_{\rm Edd}$" relation during the
rise phase, when the source luminosity is  larger than 0.03 $L_{\rm Edd}$
(dotted line in the middle panel of  Fig.~\ref{fluxnumax}; note that this
line does not  indicate a best-fit model, but is a copy of the dashed line
in the same panel, shifted by a factor of  3 along the $y-$axis).
Therefore, the $L_1$ properties (i.e. rms and the dependence of  $\nu_{\rm
max,1}$ on source luminosity), change above 0.03 $L_{\rm Edd}$ in  the rise
phase of the outburst.

The ``$\nu_{\rm max,2} -$ luminosity" relation (bottom panel in
Fig.~\ref{fluxnumax}) appears to be  roughly similar for both the rise and
the decay phases of the outburst.  The horizontal dot-dashed line in the
same panel indicates $L_{crit,S11}$, at which point  $\nu_{max,2}\sim 3$
Hz. As we discussed in the previous section, at higher frequencies  (and
hence luminosities), rms$_2$ starts decreasing. 

Since $L_1$ is present in almost all observations, and hence provides a
detailed picture of the relation between peak frequency and luminosity, we
used the best-fit  ``$\nu_{\rm max,2}-\nu_{\rm max,1}$" relation (from
section 3.1) and the best-fit  ``$\nu_{\rm max,1}-$luminosity" relation, and
found that $\nu_{\rm max,2}\propto L^{0.57}$ for  luminosities below
$L_{crit,S11}$ in the rise phase, while $\nu_{\rm max,2}\propto L^{2.3}$
when the  source luminosity is  greater than $L_{crit,S11}$ in the decay
phase. The two relations are indicated by the solid and  dashed lines in
the bottom panel of Fig.~\ref{fluxnumax}, respectively. The solid line
agrees well with the  [$\nu_{\rm max,2},(L_{\rm bol}/L_{\rm Edd})$] data in
the rise phase, all the way up to the highest luminosity at  which $L_2$ is
detected. Regarding the decay phase, $L_2$ is detected in a few PSDs when
the source luminosity  is greater than $L_{crit,S11}$. Nevertheless, the
few available points agree with the dashed line.  This suggests that  the
$L_2$ ``peak frequency -- luminosity" relation above $L_{crit,S11}$ may not
be the same  during the rise and decay phases, assuming that $L_2$ still
operates even in the cases when we cannot  detect it. 

During the decay phase, $L_2$ is not detected until $L_{\rm bol}/L_{\rm
Edd}$  decreases below $0.01L_{\rm Edd}$, while during the rise phase, we
cannot detect $L_2$ when  the source luminosity is larger than $\sim 0.03
L_{\rm Edd}$.  In both cases, $\nu_{\rm max,2}\sim 10$Hz, and rms$_{2}\sim
10$\%. In fact, the  ``$\nu_{\rm max,2}-$rms" anti-correlation (shown in
the bottom panel of Fig.~\ref{results2}) indicates  that the $L_2$ rms
should keep decreasing at higher luminosities. The combination of these two
facts  (increasing peak frequency above 10 Hz, and decreasing rms below
10\%, as the luminosity increases)  suggests that $L_2$ may be always
present, but cannot be detected because it becomes too weak and shifts
to frequencies too high to be detected in the present PSDs, especially so
if  $L_2$ were affected by a similar frequency discontinuity to that seen
in $L_1$ at MJD 53635 (dashed line in Fig.~\ref{results1}). 

In order to quantify the $L_0$ ``peak frequency -- luminosity" relation, we
found that  $\nu_{\rm max,0}\propto L^{5.1}$ according to the best-fit 
``$\nu_{\rm max,0}-\nu_{\rm max,1}$" relation (section 3.1) and the 
best-fit  ``$\nu_{\rm max,1}$ -- luminosity" relation in the decay phase above
$L_{crit,S11}$.  This relation is indicated by the dashed line in the top
panel  of Fig.~\ref{fluxnumax}. The line agrees reasonably well with the
data in the decay phase (except perhaps  during the highest luminosity
phase, when  only $L_0$ is detected in the PSD). The dotted  line in the
same panel indicates the same relation with the normalization increased by
factor of 3.  Just like with $L_1$, this line agrees well with the observed
data. 

$L_0$ ``appears" at the same luminosity levels as when $L_2$
``disappears".  The non-detection of $L_0$ at low flux may be again due to
the sensitive and limited frequency band of the PSDs. The rms$_0$ remains
rather constant  as the outburst evolves, although at a low level of $\sim
10$\%.  At the same time, just like $L_1$, $\nu_{\rm max,0}$ shifts to
lower frequencies as the source luminosity  decreases. At the lowest
luminosity, $\nu_{\rm max,0}$ would be too low to be detected in the
frequency bandpass considered. This shift toward lower frequencies would be
enhanced if $L_0$ were affected by a similar discontinuity to that observed
in $L_1$ in the ``frequency--time" relation at MJD 53635 (see
Fig.~\ref{results1}. Consequently, it is possible that $L_0$  operates at
all times, and that we cannot  detect it when $\nu_{\rm max,0}$ shifts to
frequencies lower than $\sim 0.2$ Hz, when both   $L_1$ and $L_2$ are
detected in the PSDs.  

Alternatively, $L_0$ and $L_2$ might physically disappear at lower/higher
luminosities, respectively because the source changes its spectral state.
This possibility is supported by the fact that the moments of appearing and
disappearing of $L_0$ and $L_2$ coincide with distinct spectral changes
(see dashed lines in Figs~\ref{lc} and \ref{results1}), which may indicate
the onset of the ``hard-soft-hard" transition. Thus $L_2$ would appear only
in the hard state, while $L_0$ would do so in the soft-intemediate state.

\begin{figure}
\includegraphics[width=8cm]{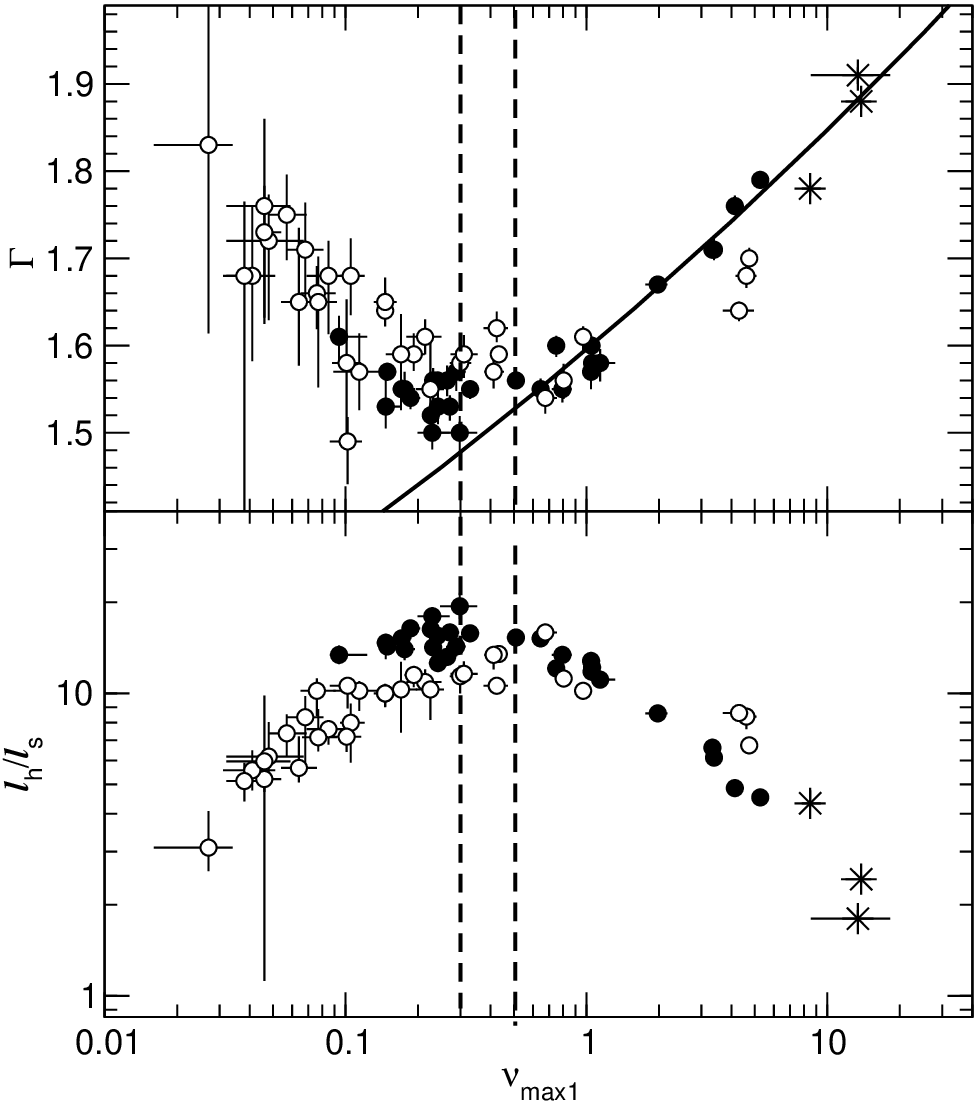} 
\caption[]{Relationship between $\nu_{\rm max,1}$ and the spectral slope
$\Gamma$ (top panel)  and the compactness ratio, $l_{\rm h}/l_{\rm s}$
(bottom panel).The dashed lines indicate the range of $\nu_{max,1}$ values
at which the  ``spectral -- timing" relation in \gro\ changes.
Filled/open circles indicate the rise and decay phase data. 
Stars represent the hard-intermediate state observations.}
\label{spectraltim}
\end{figure}

\subsection{The spectral--timing relations in \gro}

S11 fitted the hard X-ray spectrum ($E\ge 2$ keV)  of \gro\ using two
different models: 1) a thermal Comptonisation model ({\tt THCOMP}),  and 2)
a hybrid thermal/non-thermal Comptonisation model ({\tt EQPAIR}) (plus a
{\tt DISKBB}  in both cases).   The spectral shape in the first case is
determined by the photon index $\Gamma$.  In the second model,  it is the
ratio of power in the hot electrons to that in the seed photons, $\ell_{\rm
h}/\ell_{\rm s}$, that determines the shape of the spectrum. This so-called
{\em compactness} ratio depends mainly on the geometry of the accretion
flow.

We show in this work that the Lorentzian peak frequencies correlate tightly
with luminosity. Since S11 showed that both $\Gamma$ and $\ell_{\rm
h}/\ell_{\rm s}$ correlate with the source luminosity, we expect a strong
correlation between Lorentzian peak frequencies, $\Gamma$ and $\ell_{\rm
h}/\ell_{\rm s}$, as well.  This is shown in Fig.~\ref{spectraltim}. Due
to the fact that   $L_1$ is detected in most of the PSDs, in this figure we
plot  $\Gamma$ vs $\nu_{\rm max,1}$  (top panel) and the $(\ell_{\rm
h}/\ell_{\rm s}) - \nu_{\rm max,1}$ relation (bottom panel). 

Both panels in Fig.~\ref{spectraltim} show that the relationship between 
spectral shape and $\nu_{\rm max,1}$ is bimodal.  At frequencies lower
than  the frequencies indicated by the dashed lines in
Fig.~\ref{spectraltim} (when  $L\simless 0.006-0.01 L_{\rm Edd}$),
$\nu_{max,1}$ anti-correlates with spectral shape: as the peak frequency
increases (together with increasing flux) the spectrum hardens (i.e.
$\Gamma$ decreases) and, equivalently, $\ell_{\rm h}/\ell_{\rm s}$
increases.  However, at higher frequencies (i.e. luminosities) the photon
index ($\ell_{\rm h}/\ell_{\rm s}$) and  peak frequency correlate
positively (anti-correlate). 

Note that, in both panels, we do {\it not} detect a strong ``hysteresis"
effect   in the sense that, at frequencies higher than those indicated by
the vertical  dashed lines, the ``spectral -- timing" relation is more or
less the same during the  rise and decay phases of the outburst. This is
mostly due to the fact that  a ``hysteresis" effect is also observed in
both $\Gamma$ and $\ell_{\rm h}/\ell_{\rm s}$ vs $L$ plots  (cfr. Fig. 2 and
Fig. 4 in S11).  Because the effect follows the same direction in both
spectral and timing parameters when plotted as a function of luminosity,
the amplitude of the effect reduces when  $\Gamma$ or $\ell_{\rm
h}/\ell_{\rm s}$ are plotted against $\nu_{\rm max}$.

We used the ($\Gamma, \nu_{\rm max,1}$) data above $0.01L_{\rm Edd}$ during
the rise phase,  and we found that the power-law relation: $\nu_{\rm
max,1}\propto \Gamma^{0.06\pm 0.01}$,   (solid line in the top panel of
Fig.~\ref{spectraltim}) describe well the positive correlation between 
$\nu_{max,1}$ and $\Gamma$ at high luminosities. In fact, this relation
describes the correlation  for both the rise and decay data sets, in
agreement with the comments we made above about the weakness of the
hysteresis effects in these plots.

\section{Discussion}

In this work we present the results from the study of the aperiodic
variability of the black-hole binary \gro\ during the 2005 outburst. We
analysed the same observations as S11.  They include all the
observations that showed a power-law photon index below 2.  In the
classification scheme defined by the ``hardness-intensity" diagram
\citep[see e.g.][]{belloni10}, all our observations, except for those at
the beginning of the decay, correspond to the canonical hard state. The
first eight observations of the decay correspond to the hard-intermediate
state or the transition between the two states \citep{motta12}.  While S11
performed a spectral variability analysis of \gro,  we concentrate on its
broad-band noise properties and variability. We followed an approach which
is  similar to that of \citet{nowak00}, \citet{pottschmidt03}, and
\citet{axelsson05} on Cyg X--1, namely, we obtained a power spectrum for
each observation and fit it with a number of Lorentzian profiles. It has
been shown that the multi-Lorentzian fit  provides a unified description of
the timing features across all kinds of accreting X-ray binaries
\citep{belloni02,klein-wolt08,reig08}. 

We find that the power spectral density (PSD) or simply power spectra are
well fitted by  a combination of three broad Lorentzians ($L_0, L_1$ and
$L_2$) and a QPO, which we  did not consider.  At the highest flux, the
PSDs are well  fitted by a single Lorentzian ($L_0$). At all other cases,
the PSDs are well fitted by the sum of  two Lorentzians, either $L_0$ and
$L_1$ (at high flux), or $L_1$ and $L_2$ (at low flux). The  whole PSD
depends strongly on the source luminosity:  it shifts to higher
frequencies, as the luminosity increases during the rise phase of the 
outburst, and then shifts back to lower frequencies as the luminosity
decreases during the decay  phase, in agreement with what is observed in
other black-hole binaries. 

\begin{figure}
\includegraphics[width=8cm]{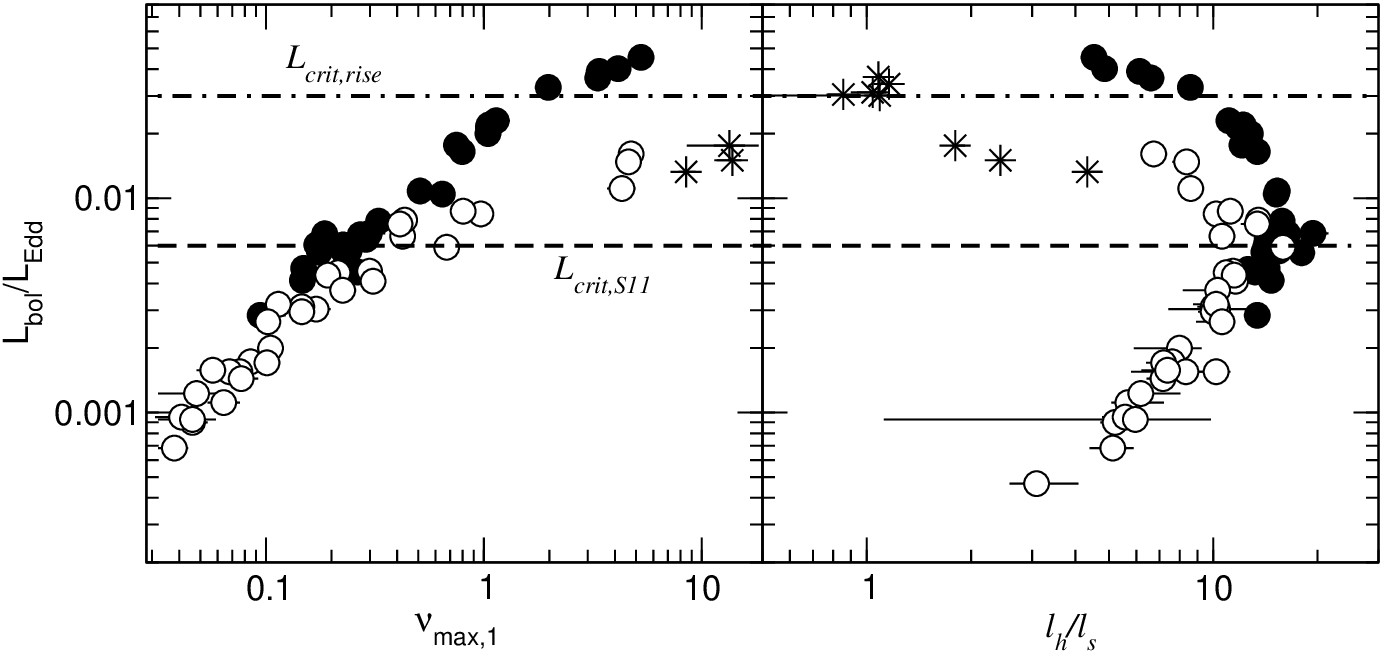} 
\caption[]{The ``$\nu_{\rm max,1}-$luminosity" relation (left panel) and
the ``$\ell_{\rm h}/\ell_{\rm s}-$luminosity" relation (right panel)  in
\gro. Filled/open circles indicate the rise and decay phase data. 
Stars represent the hard-intermediate state observations. The dashed and
dot-dashed lines indicate the $L_{crit,S11}$ ($\sim L_{crit,decay})$ and
$L_{crit,rise}$ luminosity levels, respectively.}
\label{final}
\end{figure}

Our first main result is the detection of a strong ``hysteresis" effect:
$\nu_{\rm max,0,decay}$ and  $\nu_{\rm max,1,decay}$ are always higher than
$\nu_{\rm max,0,rise}$ and  $\nu_{\rm max,1,rise}$, as long as the source
luminosity is greater than $L_{crit,S11}$. Hysteresis is a well-known
effect  observed in the spectral parameters (i.e., photon index, X-ray
colours, iron line strength) of transient   \citep{homan01,nowak02,dunn08}
and persistent \citep{smith07} black-hole binaries, and neutron-star
binaries \citep{maccarone03}. Most remarkable is the hysteresis observed in
the X-ray colours that gives rise to the characteristic $q$-shape of the
hardness-intensity diagram of black-hole transients \citep{belloni05}. A
hysteresis effect between the root mean  square (rms) amplitude of the
variability and the net count-rate has also been reported
\citep{munoz-darias11}. Even the near-IR flux seems to show a hysteresis
effect in black-hole state transitions \citep{russell07}. In this work we
show that hysteresis is also present in the evolution of the  Lorentzian
peak frequencies during the outburst of \gro\footnote{Some evidence for a
hysteresis pattern associated with a broad-band component can be seen in
Fig. 7 of \citet{belloni05}, although these authors do not discuss the
implications of such result.}. 

Our second major result is that, just like the spectral properties of the
source, the hard state timing properties of \gro\ also change above/below a
``critical" luminosity  level. However, we also find that this level is
different during the rise and the decay phase of the outburst:  $L_{crit,
rise}\sim 0.03L_{\rm Edd}$, while  $L_{crit, decay}\sim L_{crit,S11}\sim
0.006L_{\rm Edd}$.  Fig.~\ref{final} shows the ``$L_1$ peak frequency --
luminosity" diagram (based on the results presented  in this work; left
panel), and the ``$\ell_{\rm h}/\ell_{\rm s}$ -- luminosity" diagram 
(using the S11 results; right panel). The dot-dashed and dashed lines
indicate the  $L_{crit,rise}$ and $L_{crit,S11}$ levels, respectively.  The
hysteresis effect is evident in both plots, but the details differ.  Both
the rise and the decay ``spectral -- luminosity" relation change around
$L_{crit,S11}$. At luminosities greater than $L_{crit,S11}$, the rise and
decay data  follow two {\it different} paths. Moreover,  the
``$L_1$ peak frequency --luminosity"  relation changes above 
$L_{crit,rise}$ (and not $L_{crit,S11}$) during the rise phase. However,
the data above the critical levels are characterised by a similar slope (i.e.
$\nu_{\rm max,1}\propto (L_{\rm bol}/L_{\rm Edd})^4$) during both the rise
and the  decay phases.   

At lower luminosity, below $L_{crit,S11}$, we observe the same ``peak
frequency -- luminosity" relation  for both the rise and decay phases. Note
that the ``$\ell_{\rm h}/\ell_{\rm s}$ -- luminosity" diagram is analogous
to the traditional q-shaped ``hardness-intensity" diagram, where X-ray
colors have been replaced by the ratio $\ell_{\rm h}/\ell_{\rm s}$ and the
count rate by luminosity (compare the right panel in Fig.~\ref{final} with the
top panel in Fig 1 in \citealt{motta12}). The luminosity level $L_{crit,S11}$
corresponds to a 3--20 keV count rate of PCU2 $\sim 50$ count s$^{-1}$,
which is the count rate below which colours and variability levels (rms)
follow also the same trend.
The ``hardness-intensity" diagram (Fig. 1 in \citealt{motta12}), together with
Figs~\ref{spectraltim} and \ref{final} illustrate the common behaviour of
the spectral, timing and colour  parameters below
$L_{crit,S11}$, irrespective of whether the source is in the rise or decay
of the outburst.

A model that is frequently assumed to explain the X-ray properties of BHBs
during their outbursts is the so called ``truncation disc" model. According
to this model, the standard cool, optically thick,  geometrically thin
accretion disc \citep{shakura73} is truncated at some radius,  $R_{trunc}$,
which is greater than the radius of the innermost stable circular orbit
radius, $R_{in}$. A hot, optically thin, geometrically thick accretion flow
(usually refereed to as the ``corona") is assumed to extend from
$R_{trunc}$ to $R_{in}$. A fraction of the  disc photons are incident on
the hot flow where they are Compton-up-scattered by the hot electrons,
giving rise to the  hard X-ray, power-law like component in the X-ray
spectra. 

As the source luminosity increases/decreases during the rise/decay phase of
the outburst,  $R_{trunc}$ is believed to decrease/increase, respectively.
If the Lorentzian frequencies correspond to a characteristic time scale at
$R_{trunc}$ (i.e. at the outer radius of the corona),  it is natural to
expect that  the PSDs will also shift to higher/lower frequencies during
the rise/decay phase of the outburst, as observed. If this is the case, 
the hysteresis effect we observe suggests that, whatever the truncation
physical mechanism may be, it  may {\it not} operate at the same rate with
luminosity: at the {\it same} luminosity level, 
$R_{trunc,rise}>R_{trunc,decay}$ (as long as $L>L_{crit,S11}$). 

The $\nu_{\rm max,1}$ may be identified with the viscous time scale at
$R_{trunc}$,  which should modulate the propagation of fluctuations in mass
accretion rate within the hot flow. Recent calculations of the viscous
frequency, $\nu_{visc}$, as a function of radius in a BH binary predict 
values of the order of $\sim 3\times 10^{-2}-10$ Hz, when  $50R_g \ge
R_{trunc} \ge 5-8 R_g$ \citep[see the green and red lines in the middle panel
of Fig. 1 in][]{ingram12}, where  $R_g$ is the gravitational radius of the
BH. This is very similar to the range of  $\nu_{\rm max,1}$ values we
observe in \gro.  Interestingly, the same authors show that, in the case of
tilted accretion flows, the ``$\nu_{visc} -$ radius" relation should 
steepen considerably at radii smaller than the so-called ``bending wave
radius", $R_{bw}$,  which is expected from a misaligned flow.
Consequently,  if $\nu_{\rm max,1}\sim \nu_{visc}(R_{trunc})$, and 
$R_{trunc}$ decreases with increasing  source luminosity, then a steepening
of the  ``$\nu_{\rm max,1}-(L_{\rm bol}/L_{\rm Edd})$" relation (identical
during the rise and  decay phase) should appear when $R_{trunc}$ becomes
smaller than $R_{bw}$. The ``critical" luminosity  levels should correspond
to the source luminosity when $R_{trunc}<R_{bw}$, and since 
$L_{crit,rise}\ne L_{crit,decay}$,  $R_{trunc}$ should become smaller than 
$R_{bw}$ at different luminosity levels during the rise and decay phases. 

Another possibility though  is that the ``$R_{trunc}-$luminosity" relation
is the same during both the rise and  decay phases, but the bending wave
radius in the decay phase is greater than $R_{bw}$ in the  rise phase. In
this case, the steep $\nu_{visc}-R_{trunc}$ relation in the decay phase 
will last until lower luminosities, hence $L_{crit,decay}<L_{crit,rise}$,
as observed.  At the same time, since $\nu_{visc}$ depends on $R_{bw}$ 
\citep[see equation 2 in][]{ingram12},  this scenario may also explain the 
difference we observe in the normalization of the $\nu_{\rm max,1}-L_{\rm
bol}/L_{\rm Edd}$ relation in the  rise and decay phases.  Nevertheless,
even in this case, since $R_{bw}$ depends on the ratio of the height of the
flow over its  radius, our results indicate that the evolution of the
disc/coronal parameters with luminosity should not be the same during the
rise and decay phase of the outburst. 

The coronal emission may also vary due to magneto-acoustic waves
propagating within the hot flow.  \citet{cabanac10} have investigated the
timing properties of the hot corona in the case when an external 
excitation propagates radially within the corona at the sound speed. They
showed that in this case,  the filtering effect  of the corona leads to the
appearence of broad-band noise components in the PSDs, which are well
fitted  by zero-centered Lorentzians. The peak frequencies of these
Lorentzians should  scale as $\sim 2.5\times 2\pi c_s/r_j$,  where $r_j$ is
the outer radius of the corona, and $c_s$ is the sound speed. In the case
of a 10 solar mass BH, a hot corona of a temperature $\sim 100$ keV, and
$50R_g\ge r_j=R_{trunc}\ge 10 R_g$,  the Lorentzian peak frequencies should
range between $\sim 1.7-8.5$ Hz, which is almost identical to the  observed
range of $\nu_{\rm max,2}$. 

S11 observed that the ``spectral shape -- luminosity" relation in \gro\
changes when $L\sim 0.006L_{\rm Edd}$: at lower (higher) luminosities
$\Gamma$ and $L_{\rm bol}/L_{\rm Edd}$ anti-correlate (correlate
positively).  Within the ``truncation disc" model, they interpreted this 
result as an evidence for a change in the X-ray radiation mechanism,
whereby the  seed photons that feed the Compton process have a different
origin depending on luminosity. At luminosities less than $\sim 0.01 L_{\rm
Edd}$, the seed photons may come from the  cyclo-synchrotron radiation of
the hot flow itself. Above the ``critical" luminosity limit, the seed
photons could stem from the thermal emission of the truncated disc.  Within
the truncation disc model, and following S11's interpretation, the fact
that  $L_0$ is detected when $L>L_{crit,S11}$ (both during the rise and the
decay phase of the outbursts) may suggest that  $L_0$ is  associated with
variability processes of the accretion disk seed soft photons.  The seed
photon flux may vary on the viscous time scale (of the disc) at
$R_{trunc}$,  and could  induce a variation in the output of the hot
corona.  If what we see is actually the Comptonized radiation (not the
disc), even at this relatively high luminosity, then the small rms
amplitude exhibited by $L_0$ can be explained because  the temperature of
the corona can adjust very quickly to the changes of soft photon flux from
the disc. In this case, the fluctuations are damped, i.e. huge fluctuation
of soft seed photons flux leads to modest change in Comptonized flux
\citep{malzac00}. This would also be consistent with the interpretation of
S11:  at luminosities $< 0.01 L_{\rm Edd}$ synchrotron seed photons
dominate  over disc photons and the disc variability cannot be seen anymore
in the Comptonized radiation.  On the other hand, if the radiation at
$L>L_{crit,S11}$ came from the disc, then we would conclude that the disc
is not varying significantly when the source luminosity is high.


Finally, the ``spectral -- timing" correlations we observe between the
Lorentzian peak frequencies and the hard X-ray  spectral slope should be a
by-product of the individual relations between $\nu_{\rm max}$ and $\Gamma$
with  $(L_{\rm bol}/L_{\rm Edd})$ that are reported in this work and in
S11. Interestingly, the $\nu_{\rm max} \propto \Gamma^{0.06}$  relation
that we observe above $L_{crit}$ is identical to a similar relation
reported by \citet{papadakis09} for a few,  bright AGN. This is one more
evidence for the similarity between the spectral and variability properties
of BHBs and AGN. In addition, this result reinforces the interpretation of
\citet{papadakis09} that this relation in AGN is due to the fact that both 
the characteristic frequencies and the average spectral slope do depend on
the source  luminosity (and ultimately on accretion rate), hence the
presence of this relation in AGN, just like in \gro\, when the  source
luminosity exceeds the ``critical" level of $\sim 0.01$ of the Eddington
limit.

\section{Summary and conclusion}

We have studied the aperiodic variability of the black-hole binary \gro\ at
the beginning and end of its 2005 outburst. We found that, just like with
its spectral evolution, the variability  evolution of the  broad-band
noise components in the power spectrum with luminosity follows the same
trend for both the rise and decay data, as long as the source luminosity is
smaller than  $\sim 0.006$ of the Eddington limit. At higher luminosities,
the evolution  of the broad-band noise with luminosity is different
during the rise and decay phases, and a strong hysteresis pattern appears. 
Our results can be summarised as follows:

{\it The PSD evolution during the rise phase.} 1) $L\le L_{crit,rise}$: The
PSDs are well  described by the sum of $L_1$ and $L_2$.  Both $\nu_{\rm
max,1}$ and $\nu_{\rm max,2}$ shift to  higher frequencies  as the source
luminosity increases: $\nu_{\rm max,1}\propto (L_{\rm bol}/L_{\rm Edd})$ 
and $\nu_{\rm max,2}\propto (L_{\rm bol}/L_{\rm Edd})^{0.57}$. 2) $L\ge
L_{\rm crit,rise}:$ The $L_1$ rms starts decreasing with source  flux, and
$\nu_{\rm max,1}\propto (L_{\rm bol}/L_{\rm Edd})^{4}$. $L_2$  is not
detected anymore, while $L_0$ appears.  Its peak frequency also increases
with increasing  luminosity acording to the relation: $\nu_{\rm
max,0}\propto (L_{\rm bol}/L_{\rm Edd})^{5.1}$. 

{\it The PSD evolution during the decay phase.} 1) $L\ge L_{crit,decay}$:
The PSDs are well fitted mainly  by the sum of $L_0$ and $L_1$. As the
luminosity decreases, $\nu_{\rm max,0}$ and $\nu_{\rm max, 1}$ also
decrease. We find that $\nu_{\rm max,1}\propto (L_{\rm bol}/L_{\rm
Edd})^4$, and  $\nu_{\rm max,0}\propto (L_{\rm bol}/L_{\rm Edd})^{5.1}$,
similarly to what we observe in the rise phase, when $L>L_{crit,rise}$. 
The only difference is that the $\nu_{\rm max}-L$ relations have a
normalization which is $\sim 3$ times smaller than before. 2)  $L\le
L_{crit,decay}$: The $L_1$ and $L_2$ peak frequencies shift to lower
frequencies as  the luminosity decreases, in a fashion similar to the rise
phase, when the source luminosity  was smaller than $L_{crit,rise}$.

We attempt to explain our results within the truncation disc model and
conclude that each broad-band noise component may be attributed to a
different process in the accretion disc: variability processes associated
with the seed photon flux ($L_0$), viscous time scales at the truncation
radius ($L_1$), and magneto-acustic waves propagating in the accretion flow
($L_2$). Irrespective though of the physical identification of the
Lorentzian peak frequencies, our results strongly suggest that, if the
truncation disc/corona model is valid, the evolution of the disc/coronal
properties with luminosity cannot be the same during the rise and decay
phases of the outburst."

\section*{Acknowledgments}

JM acknowledges financial support from the French Research National Agency:
CHAOS project ANR-12-BS05-0009 (http://www.chaos-project.fr) and from PNHE
in France.


\begin{thebibliography}{}


\bibitem[Axelsson et al.(2005)]{axelsson05}
Axelsson, M., Borgonovo, L., \& Larsson, S., 2005, A\&A, 438, 999
\bibitem[Belloni et al.(2002)]{belloni02}
Belloni, T., Psaltis, D., \& van der Klis, M., 2002, ApJ, 572, 392 
\bibitem[Belloni et al.(2005)]{belloni05}
Belloni, T., Homan, J., Casella, P., van der Klis, M., Nespoli, E., Lewin,
W. H. G., Miller, J. M., \& Méndez, M.,  2005, A\&A, 440, 207
\bibitem[Belloni(2010)]{belloni10}
Belloni, T., 2010, LNP, 794, 53
\bibitem[Cabanac et al.(2010)]{cabanac10}
Cabanac, C., Henri, G., Petrucci, P.-O., Malzac, J., Ferreira, J., \&
Belloni, T. M., 2010, MNRAS, 404, 738
\bibitem[Casella et al.(2005)]{casella05}
Casella, P., Belloni, T., \& Stella, L, 2005, ApJ, 629, 403
\bibitem[Chakrabarti et al.(2008)]{chakrabarti08}
Chakrabarti, S. K., Debnath, D., Nandi, A., \& Pal, P. S., 2008, A\&A, 489,
L41
\bibitem[Cowley et al.(1983)]{cowley83}
Cowley, A. P., Crampton, D., Hutchings, J. B., Remillard, R., \& Penfold,
J. E., 1983, ApJ, 272, 118
\bibitem[Debnath et al.(2008)]{debnath08}
Debnath, D., Chakrabarti, S. K., Nandi, A., \& Mandal, S., 2008, BASI, 36, 151
\bibitem[Dunn et al.(2008)]{dunn08}
Dunn, R. J. H., Fender, R. P., K\"ording, E. G., Cabanac, C., \& Belloni,
T., 2008, MNRAS, 387, 545
\bibitem[Foellmi et al.(2006)]{foellmi06}
Foellmi, C., Depagne, E., Dall, T.H., \& Mirabel, I.F., 2006, A\&A, 2006,
457, 249
\bibitem[Gallo et al.(2008)]{gallo08}
Gallo, E., Homan, J., Jonker, Peter G., \& Tomsick, J. A., 2008, ApJ, 683, L51
\bibitem[Gierlinski et al.(2001)]{gierlinski01}
Gierli\'nski, M., Maciolek-Nied\'zwiecki, A., \& Ebisawa, K., 2001, MNRAS,
325, 1253
\bibitem[Hjellming \& Rupen(1995)]{hjellming95}
Hjellming, R. M., \& Rupen, M. P., 1995, Nature 375, 464
\bibitem[Homan et al.(2001)]{homan01}
Homan, J., Wijnands, R., van der Klis, M., Belloni, T., van Paradijs, J.,
Klein-Wolt, M., Fender, R., \& M\'endez, M., 2001, pJS, 132, 377
\bibitem[Ingram \& Done(2012)]{ingram12}
Ingram, A., \& Done, C., 2012, MNRAS, 419,2369
\bibitem[Jahoda et al.(1996)]{jahoda96}
Jahoda, K., Swank, J.H., Giles, A.B., Stark, M.J., Strohmayer, T.,
Zhang, W., \& Morgan, E.H.,  1996, SPIE, 2808, 59
\bibitem[Klein-Wolt \& van der Klis(2008)]{klein-wolt08}
Klein-Wolt, M., \& van der Klis, M., 2008, ApJ, 675, 1407
\bibitem[Kylafis et al.(2008)]{kylafis08}
Kylafis, N. D., Papadakis, I. E., Reig, P., Giannios, D., \& Pooley, G. G.,
2008, A\&A, 489, 481
\bibitem[Maccarone \& Coppi(2003)]{maccarone03}
Maccarone, T. J., \& Coppi, P. S. 2003, MNRAS, 338, 189
\bibitem[Malzac \& Jourdain(2000)]{malzac00}
Malzac, J., \& Jourdain, E.,  2000, A\&A, 359, 843
\bibitem[Markoff et al.(2005)]{markoff05}
Markoff, S., Nowak, M. A., \& Wilms, J.,  2005, ApJ, 635, 1203
\bibitem[Motta et al.(2012)]{motta12}
Motta, S., Homan, J., Mu\~noz Darias, T., Casella, P., Belloni, T.
M., Hiemstra, B., \& M\'endez, M.,  2012, MNRAS, 427, 595
\bibitem[Mu\~noz-Darias et al.(2011)]{munoz-darias11}
Mu\~noz-Darias, T., Motta, S., \& Belloni, T. M., 2011, MNRAS, 410, 679
\bibitem[Nandi et al.(2012)]{nandi12}
Nandi, A., Debnath, D., Mandal, S., \& Chakrabarti, S. K., 2012, A\&A, 542,
A56
\bibitem[Nowak et al.(1999)]{nowak99}
Nowak, M. A., Vaughan, B. A., Wilms, J., Dove, J. B., Begelman, M. C., 
1999, ApJ, 510, 874
\bibitem[Nowak(2000)]{nowak00}
Nowak, M. A., 2000, MNRAS, 318, 361
\bibitem[Nowak et al.(2002)]{nowak02}
Nowak, M. A., Wilms, J., Dove, J. B.,  2002, MNRAS, 332, 856
\bibitem[Orosz \& Bailyn(1997)]{orosz97}
Orosz, J. A., \& Bailyn, C. D., 1997, ApJ, 477, 876
\bibitem[Papadakis et al.(2009)]{papadakis09}
Papadakis, I. E., Sobolewska, M., Arevalo, P., Markowitz, A., McHardy, I.
M., Miller, L., Reeves, J. N., \& Turner, T. J., 2009, A\&A, 494, 905
\bibitem[Pottschmidt et al.(2003)]{pottschmidt03}
Pottschmidt, K., Wilms, J., Nowak, M. A., Pooley, G. G., Gleissner, T., 
Heindl, W. A., Smith, D. M., Remillard, R., \& Staubert, R., 2003, A\&A,
407, 1039
\bibitem[Press et al.(1992)]{press92}
Press, W. H., Teukolsky, S. A., Vetterling W. T., \& Flannery , B. P.,
1992, in {\it Numerical Recipes: The
art of scientific computing}, Cambridge University Press.
\bibitem[Reig(2008)]{reig08}
Reig, P., 2008, A\&A, 489, 725
\bibitem[Remillard \& McClintock(2006)]{remillard06}
Remillard, R. A., \& McClintock, J. E., 2006, ARA\&A, 44, 49
\bibitem[Russell et al.(2007)]{russell07}
Russell, D.~M., Maccarone, T.~J., Körding, E.~G., \& Homan, J.,  2007,
MNRAS, 379, 1401
\bibitem[Shakura \& Sunyaev(1973)]{shakura73}
Shakura, N. I. \& Sunyaev, R. A., 1973, A\&A, 24, 337
\bibitem[Shaposhnikov et al(2007)]{shaposhnikov07}
Shaposhnikov, N., Swank, J., Shrader, C. R., Rupen, M., Beckmann, V.,
Markwardt, C. B., \& Smith, D. A., 2007, ApJ, 655, 434
\bibitem[Shaposhnikov \& Titarchuk(2009)]{shaposhnikov09}
Shaposhnikov, N. \& Titarchuk, L., 2009, ApJ, 699, 453
\bibitem[Smith et al.(2007)]{smith07}
Smith, D.M., Dawson, D.M., \& Swank, J.H., 2007, ApJ, 669, 1138
\bibitem[Sobolewska et al.(2009)]{sobolewska09}
Sobolewska, M. A., Gierli\'nski, M., \& Siemiginowska, A., 2009, MNRAS,
394, 1640
\bibitem[Sobolewska et al.(2011)]{sobolewska11}
Sobolewska, M. A., Papadakis, I. E., Done, C., \& Malzac, J., 2011, MNRAS,
417, 280
\bibitem[Webster \& Murdin(1972)]{webster72}
Webster, B. L., \& Murdin, P., 1972, Nature, 235, 37
\bibitem[Wu \& Gu(2008)]{wu08}
Wu, Q., \& Gu, M., 2008, ApJ, 682, 212
\bibitem[Zhang et al.(1994)]{zhang94}
Zhang, S.N., Wilson, C.A., Harmon, B. A., Fishman, G. J.,  
Wilson, R. B., Paciesas, W. S., Scott, M., \& Rubin, B. C., 1994, IAUC 6046

\end{thebibliography}
\end{document}